\documentclass[pra,twocolumn,showpacs,floatfix,superscriptaddress,aps]{revtex4}
\usepackage{amsmath}
\usepackage{makeidx}
\usepackage{amssymb}
\usepackage{graphicx}

\usepackage[usenames]{color}
\usepackage[normalem]{ulem}

\usepackage{hyperref}

\begin{document}

\title{Phase separation in a spin-orbit coupled Bose-Einstein condensate}

\author{Sandeep Gautam}
\affiliation{Instituto de F\'{\i}sica Te\'orica, UNESP - Universidade Estadual
             Paulista, \\ 01.140-070 S\~ao Paulo, S\~ao Paulo, Brazil}
             
\author{S. K. Adhikari}
\affiliation{Instituto de F\'{\i}sica Te\'orica, UNESP - Universidade Estadual
             Paulista, \\ 01.140-070 S\~ao Paulo, S\~ao Paulo, Brazil}

\date{\today}
\begin{abstract}
 We study a spin-orbit (SO) coupled hyperfine spin-$1$ Bose-Einstein 
 condensate (BEC) in a quasi-one-dimensional trap. For a SO-coupled BEC in a 
 one-dimensional box, we show that in the absence of the Rabi term, any 
 non-zero value of SO coupling will result in a phase separation among the 
 components for a ferromagnetic BEC, like $^{87}$Rb. On the other hand, SO 
 coupling favors miscibility in a polar BEC, like $^{23}$Na. In the presence 
 of a harmonic trap, which favors miscibility, a ferromagnetic BEC phase 
 separates, provided the SO-coupling strength and number of atoms are greater 
 than some critical value. The Rabi term favors miscibility irrespective of 
 the nature of the spin interaction: ferromagnetic or polar.
 \end{abstract}

\maketitle
\section{Introduction}
A Bose-Einstein condensate (BEC) with spin degrees of freedom, known as a
spinor BEC, was first experimentally realized and studied in a gas of $^{23}$Na
atoms, with hyperfine spin $F=1$, in an optical dipole trap \cite{Stamper-Kurn}. 
{ This has lead to a flurry of investigation on both the 
theoretical and experimental fronts, which has been reviewed in Ref. 
\cite{ueda}.} In the present work, we study the ground state structure of the 
$F=1$ spin-orbit (SO) coupled spinor BEC in a quasi-one-dimensional (quasi-1D) 
trap \cite{luca} within the framework of the mean-field theory. The mean-field 
theory to study $F=1$ spinor  BECs  was developed 
independently by Ohmi {\em et al.} \cite{Ohmi} and Ho \cite{Ho}. The 
SO-interaction is absent in neutral atoms and an engineering with an 
external electromagnetic field is needed for its experimental realization. 
A variety of SO couplings can be engineered by counter propagating Raman lasers 
coupling the hyperfine states, and the parameters of this coupling can be 
controlled independently \cite{young}. The SO interaction  has been  achieved 
recently with equal strengths of Rashba \cite{Bychkov} and Dresselhaus 
\cite{Dresselhaus,Liu} couplings employing a necessary engineering to obtain 
experimentally a SO-coupled BEC of two of the existing three hyperfine spin 
components of the $F=1$ state of $^{87}$Rb \cite{Lin,Linx} forming a pseudo-spinor 
BEC. This has been followed by other experiments on SO-coupled pseudo-spinor BECs 
\cite{JY_Zhang}. In the case of a $F=1$ spinor BEC, there are theoretical 
proposals to realize SO-coupling interaction involving the three hyperfine 
spin components \cite{Juzeliunas}. SO-coupled degenerate Fermi 
gases ($^{40}$K and $^{6}$Li) have also been experimentally realized 
\cite{P_Wang}. A mean-field Gross-Pitaevskii (GP) equation for the theoretical 
study of  dynamics in SO-coupled BECs has also been proposed 
\cite{Juzeliunas,ueda,H_Wang,Bao}.

The ground states of the   SO-coupled two-component 
pseudo-{ spinor} BEC  and of the three-component spinor 
BEC {  have been} theoretically investigated by Wang {\em et al.} 
\cite{Wang}. It has further been established that the SO-coupled spin-$1/2$ 
(pseudo-spinor), $F=1$ and $F=2$ spinor BECs in quasi-two-dimensional (quasi-2D) 
traps \cite{luca} can have a variety of nontrivial ground state structures 
\cite{Gopalakrishnan, Xu-1,Ruokokoski}. There have been studies of intrinsic 
spin-Hall effect \cite{Larson}, chiral confinement \cite{Merkl}, superfluidity 
\cite{super}, Josephson oscillation \cite{josep}, vortices \cite{vor}, and 
solitons \cite{sol} in a SO-coupled BECs. In general, for experimentally 
feasible parameters, the ground state of a $F=1$ spinor BEC can host a single 
vortex or a square vortex lattice for weak and strong SO coupling, respectively 
\cite{Ruokokoski}. Additionally, plane and standing wave states appear as ground 
states in the case of ferromagnetic and polar (antiferromagnetic) BECs, 
respectively, for medium strengths of SO coupling \cite{Ruokokoski}. The ground 
state of the $F = 1$ spinor BEC in the presence of a homogeneous magnetic 
field has also been studied ~\cite{Zhou,  Matuszewski-1, Matuszewski-2}.
It was shown in Refs.~\cite{Matuszewski-1,Matuszewski-2} that a uniform 
magnetic field can lead to a phase separation in polar BEC. Phase separation 
has already been observed in a pseudo-spinor BEC consisting of two hyperfine 
states of $^{87}$Rb in quasi-2D geometries \cite{Lin}. 

In this paper, we investigate the ground state of a SO-coupled $F=1$ spinor 
BEC in a quasi-1D trap. For the model of SO-coupling employed in this work, 
we find that compared to the homogeneous magnetic field, SO coupling leads 
to a phase separation in the case of a ferromagnetic BEC, whereas in the 
case of a polar BEC, it makes the miscible profile energetically more stable. 
Here, we use a numerical solution of the  generalized mean-field GP 
equation \cite{H_Wang,Bao} to study the possible phase separation between the 
different hyperfine spin components of a SO-coupled BEC. We also study the 
possibility of a phase separation in a uniform spinor condensate in a 1D box 
employing an analytical model. The results of this analytical study provide a 
qualitative understanding of the numerical findings for a trapped SO-coupled BEC.

The paper is organized as follows. In Sec.~\ref{Sec-II}, we describe the 
coupled GP equation  used to study the SO-coupled $F=1$ spinor BEC in a 
quasi-1D trap. This is followed by an analytical investigation of a SO-coupled 
spinor BEC in a 1D box in Sec.~\ref{Sec-III}. By comparing the energies of 
various competing geometries for both ferromagnetic and polar BECs, the ground 
state structure is determined from a minimization of energy. In the case of a 
mixture of two scalar BECs, similar analysis leads to the criterion for a phase 
separation \cite{Ao}. In Sec.~\ref{Sec-IV}, we numerically study the SO-coupled 
spinor BEC in a quasi-1D trap. We conclude the manuscript by providing a 
summary of this study in Sec. \ref{Sec-V}.


\section{Mean-field model for a SO-coupled BEC}
\label{Sec-II}
For the electronic states of a hydrogen-like atom the SO contribution to the 
atomic spectrum naturally appears because of the magnetic energy of this 
coupling existing due to electronic charge. In the case of the hyperfine states 
of neutral atoms, an engineering with external electromagnetic fields is 
required for the SO coupling to contribute to the BEC. We use the SO-coupled 
interaction of the experiment of  Lin {\it et al.} \cite{Lin} 
for two hyperspin components of the $^{87}$Rb hyperfine state 5S$_{1/2}$, 
realized with strength $\gamma$ 
using two counterpropagating Raman lasers of wavelength $\lambda_r$ oriented 
at an angle $\beta_r$:  $\gamma = \hbar k_r/m$, where 
$k_r = 2\pi\sin(\beta_r/2)/\lambda_r$ and $m$ is the mass of an atom. 
This SO coupling is equivalent to that of an 
electronic system with equal contribution of Rashba \cite{Bychkov} and 
Dresselhaus \cite{Dresselhaus} couplings and with an external uniform magnetic 
field.
However, here we consider the SO coupling among
the three spin components of the $F=1$ state, e.g.,  
$| F=1, m_F=1\rangle$, $| F=1, m_F=0\rangle$, and $| F=1, m_F=-1\rangle$,
where   $m_F$ is the $z$ projection of $F$. It has been shown \cite{lan}
that this SO coupling among the three
hyperfine spin components 
can be 
generated by an engineering as in Ref. \cite{Lin}.
We will consider the three 
spin components of the $F=1$ hyperfine  state 5S$_{1/2}$ of $^{87}$Rb and 
3S$_{1/2}$ of $^{23}$Na.

We consider such a  quasi-1D  hyperfine  spin-$1$ SO-coupled spinor BEC  
confined  along the $x$-axis obtained by making the trap along $y$ and $z$ 
axes much stronger than  that along the $x$-axis. The transverse dynamics of 
the BEC is assumed to be frozen to the respective ground states of harmonic
traps. Then, the single-particle quasi-1D Hamiltonian of the system under the 
action of a strong transverse trap of angular frequencies $\omega_y$ and 
$\omega_z$ along $y$ and $z$ directions respectively, can be written as 
\cite{Lin,Y_Zhang}
\begin{equation}\label{1x}
 H_0 = \frac{p_x^2}{2m}+\gamma p_x \Sigma_z + \Omega \Sigma_x,
\end{equation} 
where $p_x=-i\hbar \partial_x$ is the momentum operator along $x$ axis,
  $\Omega$ is the Rabi frequency 
\cite{Lin,Linx}, and $\Sigma_z$ and $\Sigma_x$ are the matrix representations 
of the $z$ and $x$ components of the spin-1 angular momentum operator, 
respectively, and are given by 
\begin{eqnarray}
\Sigma_z=  \left( \begin{array}{ccc}
1 & 0 & 0 \\
0& 0& 0 \\
0& 0& -1\end{array} \right), \quad 
\Sigma_x= \frac{1}{\sqrt 2} \left( \begin{array}
 {ccc}
0& 1 & 0 \\
1 & 0 & 1 \\
0 & 1 & 0\end{array} \right) .
\end{eqnarray}

If the interactions among the atoms in the BEC are taken into account, in the 
Hartree approximation, using the single particle model Hamiltonian (\ref{1x}), a 
quasi-1D \cite{luca} spinor BEC can be described by the following set of three 
coupled mean-field  partial differential GP equations for the wave-function 
components  $\psi_j$  \cite{H_Wang,Bao,ueda}
\begin{eqnarray}
 i\hbar\frac{\partial \psi_1}{\partial t} &=& 
 \left( -\frac{\hbar^2}{2m}\frac{\partial^2}{\partial x^2}
 +V(x)+c_0\rho\right)\psi_1 +c_2(\rho_1+\rho_0- \nonumber\\
  &&\rho_{-1})\psi_1+c_2\psi_{-1}^*\psi_0^2-i\hbar\gamma
  \frac{\partial\psi_1}{\partial x} + \frac{\Omega}{\sqrt{2}}\psi_0,
  \label{gp-1}\\
 i\hbar\frac{\partial \psi_0}{\partial t} &=& 
  \left( -\frac{\hbar^2}{2m}\frac{\partial^2}{\partial x^2} 
  +V(x)+c_0\rho\right)\psi_0+c_2(\rho_1+\rho_{-1})\nonumber\\
  &&\psi_0+2c_2\psi_1\psi_{-1}\psi_0^*+
  \frac{\Omega}{\sqrt{2}}(\psi_1+\psi_{-1}),\label{gp-2}\\
 i\hbar\frac{\partial \psi_{-1}}{\partial t} &=& 
 \left( -\frac{\hbar^2}{2m}\frac{\partial^2}{\partial x^2} 
 +V(x)+c_0\rho\right)\psi_{-1}+c_2(\rho_0+\rho_{-1}\nonumber\\
 &&-\rho_{1})\psi_{-1}+c_2\psi_{1}^*\psi_0^2
 +i\hbar\gamma\frac{\partial\psi_{-1}}{\partial x} + 
 \frac{\Omega}{\sqrt{2}}\psi_0\label{gp-3},
\end{eqnarray}
where $V(x) = m\omega_x^2x^2/2$ is the 1D harmonic trap, 
$c_0 = 2\hbar^2 (a_0+2a_2)/(3m l_{yz}^2)$, 
$c_2 = 2\hbar^2 (a_2-a_0)/(3m l_{yz}^2)$,  
$a_0$ and $a_2$ are the $s$-wave scattering lengths in the total 
spin $0$ and $2$ channels, respectively,   
$\rho_j = |\psi_j|^2$ with $j=1,0,-1$ are the component densities, 
$\rho = \sum_{j=-1}^1|\psi_j|^2$ is the total density, 
and $l_{yz} = \sqrt{\hbar/(m\omega_{yz})}$ with  $ \omega_{yz}=\sqrt{\omega_y\omega_z}$
is the oscillator length in the transverse $y-z$ plane. 
The normalization condition is
\begin{equation}
\int_{-\infty}^\infty dx \sum_{j=-1}^1|\psi_j(x)|^2=N.
\label{norm}
\end{equation}
In order to transform  Eqs. (\ref{gp-1}) - (\ref{gp-3}) into dimensionless 
form, we use the scaled variables defined as
\begin{equation}
 \tilde{t} = \omega_x t,~\tilde{x} = \frac{x}{l_0},
 ~\phi_j(\tilde{x},\tilde{t}) = 
 \frac{\sqrt{l_0}}{\sqrt{N}}\psi_j(\tilde{x},\tilde{t}), 
\end{equation}
where $l_0=\sqrt{\hbar/(m\omega_{x}}$) is the oscillator length along $x$-axis, 
$N$ is the total number of the atoms. Using these dimensionless variables, 
the coupled mean-field Eqs.~(\ref{gp-1}) - (\ref{gp-3}) in dimensionless form 
are
\begin{eqnarray}\label{xxa}
 i \frac{\partial \phi_1}{\partial \tilde{t}} &=& 
 \left( -\frac{1}{2}\frac{\partial^2}{\partial \tilde{x}^2}
 +\tilde{V}+\tilde{c}_0\tilde{\rho}\right)\phi_1
 +\tilde{c}_2(\tilde{\rho}_1+\tilde{\rho}_0\nonumber \\
  &&-\tilde{\rho}_{-1})\phi_1+\tilde{c}_2\phi_{-1}^*\phi_0^2
  -i \tilde{\gamma}\frac{\partial\phi_1}{\partial \tilde{x}} 
 + \frac{\tilde{\Omega}}{\sqrt{2}}\phi_0,\label{gp_s1}\\
 i \frac{\partial \phi_0}{\partial \tilde{t}} &=& 
 \left( -\frac{1}{2}\frac{\partial^2}{\partial \tilde{x}^2} 
  +\tilde{V}+\tilde{c}_0\tilde{\rho}\right)\phi_0+
 \tilde{c}_2(\tilde{\rho}_1+\tilde{\rho}_{-1})\nonumber\\
  &&\phi_0+2\tilde{c}_2\phi_1\phi_{-1}\phi_0^*
   +\frac{\tilde{\Omega}}{\sqrt{2}}(\phi_1+\phi_{-1}),\label{gp_s2}\\
 i \frac{\partial \phi_{-1}}{\partial \tilde{t}} &=& 
 \left( -\frac{1}{2}\frac{\partial^2}{\partial \tilde{x}^2} 
 +\tilde{V}+\tilde{c}_0\tilde{\rho}\right)\phi_{-1}
 +\tilde{c}_2(\tilde{\rho}_0+\tilde{\rho}_{-1}\nonumber\\
 &&-\tilde{\rho}_{1})\phi_{-1}+\tilde{c}_2\phi_{1}^*\phi_0^2
 +i \tilde{\gamma}\frac{\partial\phi_{-1}}{\partial \tilde{x}} 
 + \frac{\tilde{\Omega}}{\sqrt{2}}\phi_0\label{gp_s3},\label{xxc}
\end{eqnarray}
where $\tilde{V} = \tilde{x}^2/2$, 
$\tilde{\gamma} = \hbar k_r/(m\omega_x l_0)$, 
$\tilde{\Omega} = \Omega/(\hbar \omega_x)$,
$\tilde{c}_0 = 2N (a_0+2a_2)/(3l^2 _{yz})$, 
$\tilde{c}_2 = 2N (a_2-a_0)/(3l^2_{yz})$,
$\tilde{\rho}_j = |\phi_j|^2$ with $j=1,0,-1$, 
and $\tilde{\rho} = \sum_{j=-1}^1|\phi_j|^2$.
The normalization condition satisfied by $\phi_j$'s is
\begin{equation}
 \int_{-\infty}^{\infty} \sum_{j=-1}^1\tilde{\rho}_j(\tilde{x})d\tilde{x} = 1.
\end{equation}
Another useful quantity $-$ magnetization $-$ related to the component 
densities is defined by 
\begin{eqnarray}
 {\cal M} = \int_{-\infty}^{\infty}[\tilde{\rho}_1(\tilde x) 
 -\tilde{\rho}_{-1}(\tilde x)]d\tilde x.
\label{magnetization}
\end{eqnarray}
Depending on the value of $\tilde c_2$ ($>0$ or $<0$) the system develops 
interesting physical properties. The interaction in the $5$S$_{1/2}$ state of 
$^{87}$Rb with $\tilde c_2<0$ is termed ferromagnetic and that in the 
$3$S$_{1/2}$ state of $^{23}$Na with $\tilde c_2>0$ is termed antiferromagnetic 
or polar. For the sake of simplicity of notations, we will represent the 
dimensionless variables without tilde in the rest of the paper.

The energy of the spinor BEC in the presence of a SO coupling is 
\cite{Bao,H_Wang}
\begin{eqnarray}
 E &=&  N\int_{-\infty}^{\infty} \Bigg\{\frac{1}{2}\left|\frac{d\phi_1}{dx}
  \right|^2+\frac{1}{2}\left|\frac{d\phi_0}{dx}\right|^2+
  \frac{1}{2}\left|\frac{d\phi_{-1}}{dx}\right|^2+V \rho\nonumber\\
  &&+ \frac{c_0}{2}\rho^2+ \frac{c_2}{2}\left(\rho_1 + \rho_0-\rho_{-1} 
  \right)\rho_1+ 
  \frac{c_2}{2}\left(\rho_1 + \rho_{-1} \right)\rho_0\nonumber\\
  & &+  \frac{c_2}{2}\left(\rho_0 + \rho_{-1}-\rho_1 \right)\rho_{-1}+
  c_2\left[ \phi_{-1}^*\phi_0^2\phi_1^*\right.\nonumber\\
  &&\left.+\phi_{-1}(\phi_{0}^*)^2\phi_1\right]
  +\gamma\left( -i\phi_1^*\frac{d\phi_1}{dx} 
  + i\phi_{-1}^*\frac{d\phi_{-1}}{dx}\right)\nonumber\\
  & &+\frac{\Omega}{\sqrt{2}}\left(\phi_1^*\phi_0+\phi_0^*\phi_1
  +\phi_{-1}^*\phi_0+\phi_0^*\phi_{-1}\right)\Bigg\}dx.
  \label{energy}
\end{eqnarray}
Based on the form of this energy functional a few inferences can be easily 
drawn about the phase separation among the various components of a  
spinor BEC with SO coupling. The energy term proportional to $c_0$ can never 
lead to a phase separation as it contains terms 
$Nc_0\int (\rho_j^2/2+\rho_j\rho_{j'})dx$, where $j,j' = 1,0,-1$ and $j\ne j'$, and 
hence corresponds to a scenario where inter- and intra-species interactions 
are of equal strengths. The situation is analogous to a binary BEC with 
$a_{12}^2 = a_{11}a_{22}$, where 
$a_{11}$ and $a_{22}$ are intra-species and $a_{12}$ the inter-species 
scattering lengths. Such a binary BEC has equal strengths of inter- and 
intra-species nonlinearities and is always miscible in the presence of a $1$D 
harmonic trap \cite{Ao,Gautam}. Let us now look at the terms proportional to 
$c_2$. For the stable solution, the phases of the three components, 
say $\theta_j$'s with $j=-1,0,1$, should satisfy
\begin{equation}
 \theta_1+\theta_{-1} + s\pi= 2\theta_0,
\end{equation}
where $s$ is an integer \cite{Matuszewski-2,Isoshima}.
Assuming that $\theta_0 = 0$ and $s = 0$, the interaction energy part of the 
total energy (\ref{energy}) can be written as
\begin{align}\label{xxx}
 E_{\rm int} & =  N \int_{-\infty}^{\infty}\biggr\{\frac{c_0}{2}\rho^2+
  \frac{c_2}{2}\left( \rho_1^2+\rho_{-1}^2+2\rho_1\rho_0
  +2\rho_0\rho_{-1}\right.\nonumber\\
  &\left.-2\rho_1\rho_{-1}+4\sqrt{\rho_1\rho_{-1}}\rho_0\right)\biggr\}dx.
\end{align}
The system will naturally move to  a state of minimum energy, which could have  
a phase-separated or an overlapping configuration. A consideration of 
minimization of energy could reveal whether the system will prefer a ground 
state with an overlapping or a phase-separated profile.
 
It is evident from Eq.~(\ref{xxx}) that in the case of a  {\em ferromagnetic} 
BEC ($c_2<0$), there is only one term $N\int|c_2|\rho_1\rho_{-1}dx$ with 
positive energy contribution representing inter-species repulsion, which will
favor a phase separation between components $1$ and $-1$. The minimum 
contribution from this term can be zero, 
{when components $1$ and $-1$ are fully phase-separated}, 
whereas, for the rest of the $c_2$ dependent terms in $E_{\rm int}$, the 
contribution is always less than zero representing inter-species attraction. 
A maximum of overlap between the components will reduce the contribution of 
these terms to energy. Hence these terms will inhibit a phase separation.
So, the phase separation in a ferromagnetic BEC, if ever it occurs, can only 
take place between components $1$ and $-1$.   

On the other hand in the case of a {\em polar or antiferromagnetic} BEC 
($c_2>0$), all the terms in Eq. (\ref{xxx}) except $-N\int c_2\rho_1\rho_{-1}dx$ 
contribute positive energy representing inter-species repulsion. For an 
arbitrary value of magnetization $\cal{M}$, the interaction
energy can be minimized in two ways. First, by making $\rho_0=0$ 
and ensuring the maximum overlap between components
$1$ and $-1$; and, secondly, by fully phase-separating the $0$th component from the 
maximally overlapping $1$ and $-1$ components. The interaction energy in both
the cases becomes
\begin{equation}
 E_{\rm int}  =  N \int_{-\infty}^{\infty}\Bigg\{\frac{c_0}{2}\rho^2
                 +\frac{c_2}{2}\left( \rho_1^2+\rho_{-1}^2
                 -2\rho_1\rho_{-1}\right)\Bigg\}dx.
\end{equation}
Hence, the phase separation in a polar BEC, if it ever occurs, is most likely  
to take place between the $0$th component and overlapping
 $1$ and $-1$ components.

\section{SO-coupled BEC in a 1D box}
\label{Sec-III}

To understand the role of the different terms in the expression for the 
interaction energy (\ref{xxx}) on phase separation, we study an analytic model 
of a uniform (trapless) spinor BEC in a 1D box of length $2L$ localized in the 
region $-L<x<L$. In order to clearly establish the role of the different terms 
in  $E_{\rm int}$ in determining the ground state structure of the $F=1$ spinor 
BEC, first we consider the one with zero magnetization ($\cal M =$ 0).

We consider the miscible and immiscible profiles in the case of a ferromagnetic 
BEC ($c_2<0$). In the miscible case, the densities are uniform and written as 
$ \rho_j(x)\equiv n_j$. Because of the symmetry between $j=1$ and $j=-1$, it is 
natural to take $n_1 =n_{-1}$. Then the densities of the three components can be 
written as 
\begin{eqnarray}\label{ma}
  \rho_1(x)  &=& n_1, \quad \quad  \quad  \quad -L<x<L,\\
   \rho_0(x)  &=&  n_0,  \quad \quad \quad \quad -L<x<L,\\
  \rho_{-1}(x)  &=& n_{-1}=n_1, \quad   -L<x<L. \label{mc}
\end{eqnarray}
All densities are zero for $|x|\ge L$. This is the general density distribution 
for a miscible configuration which we will use in this study. In the absence of 
a SO coupling and Rabi term $(\gamma=\Omega=0)$, the interaction energy
(\ref{xxx}) for a ferromagnetic BEC in the 1D box becomes 
\begin{eqnarray}\label{enx}
 E_{\rm int} &=& NL[c_0(4n_1^2+n_0^2+4n_1n_0)-|c_2|8n_1n_0],
\end{eqnarray}
and the corresponding  normalization condition is
\begin{equation}
 2L(2n_1+n_0) = 1.
 \label{norm1}
\end{equation}
In the trapped case, as considered in Sec. \ref{Sec-II}, the energies 
(\ref{energy}) or (\ref{xxx}) are extensive properties and increase with the size of the system.
However,  the energy density (energy per unit length)  of a uniform gas, as considered in this section, is an intensive 
property \cite{Ao}     
and does not depend on system size or the total length of the box, provided that a constant particle density is maintained when the size is changed. Recalling that the constants 
$c_0$ and $c_2$ are proportional to the number of atoms $N$, Eq. (\ref{enx}), and all other  energies  in this section reveal the interesting feature
\begin{equation}
\frac{ E_{\rm int}}{L} \sim \left(\frac{N}{L} \right)^2,
    \end{equation}
also valid for nonspinor systems \cite{Ao}. 
The minimum of  energy (\ref{enx}), subject to the normalization constraint 
(\ref{norm1}) and for $n_j$'s $\ge0$, occurs at
\begin{eqnarray}
n_1  = n_{-1}' = \frac{1}{8L},~n_{0}  = \frac{1}{4L},
\label{den_mis_ferro}
\end{eqnarray}
and the corresponding minimum energy is 
\begin{equation}\label{emin}
E_{\rm int}^{\rm min(M)} =  N\frac{(c_0-|c_2|)}{4L}.
\end{equation}

In the immiscible case, where components  $j=1$ and $j=-1$ are separated, 
let $n_1'$ be the density of component $1$ from $-L$ to $0$ and 
$n_{-1}' = n_1'$ be the density of component $-1$ from $0$ to $L$.
This symmetric distribution is consistent with the symmetry between 
components $j=1$ and $-1$ in the mean-field Eqs. (\ref{xxa})-(\ref{xxc}).
The density of component $0$ distributed from $-L$ to $L$ is taken to be 
$n_{0}'$ as in the  miscible case, so that, 
\begin{eqnarray}\label{xa}
  \rho_1(x)  &=& \left\{
     \begin{array}{lr}
        n_1', &  \quad  \quad\quad \quad  -L<x<0,\\
        0, & \quad \quad \quad \quad  ~0\leq x\leq L,
     \end{array}
   \right.\\
   \rho_0(x)  &=& 
     \begin{array}{lr}
        n_0', & \quad \quad \quad \quad  \quad   -L<x<L,\\
     \end{array}
\\
   \rho_{-1}(x)  &=& \left\{
     \begin{array}{lr}
        n_{-1}'=n_1', &  ~0<x<L,\\
        0, &  ~-L\leq x\leq0.
     \end{array}
   \right.\label{xc}
\end{eqnarray}
All densities are zero for $|x|\ge L$. This is the general density distribution 
for an immiscible configuration, which we will use in this study for the
ferromagnetic condensate. As mentioned in Sec. \ref{Sec-II}, for a 
ferromagnetic BEC, a phase separation between the $1$ 
and $-1$ components is energetically the most favorable among all other 
possible phase separations. This is the reason to choose the aforementioned 
distribution for the immiscible profile. The interaction energy for this 
distribution is
\begin{eqnarray}
 E_{\rm int} = NL[c_0(n_1'^2+n_0'^2+2n_1'n_0')
              -|c_2|n_1'(n_1'+2n_0')],
 \end{eqnarray}
with  the normalization condition 
 \begin{equation}
  2L(n_1'+n_0') =1.
  \label{norm2}
 \end{equation}
The condition of the minimum of energy in this case, again subject to the 
normalization constraint  ~(\ref{norm2}) and for  $n_j$'s $\ge0$, is
\begin{eqnarray}
n_1'  =  n_{-1}' = \frac{1}{2L},~n_0'  =  0,
\end{eqnarray}
and the minimum value of interaction energy $ E_{\rm int}^{\rm min(I)}$ is
the same as in the miscible case, given by Eq. (\ref{emin}): 
$E_{\rm int}^{\rm min(M)}= E_{\rm int}^{\rm min(I)}$. 
Thus, from an energetic consideration, the miscible and immiscible profiles 
are equally favorable in a homogeneous ferromagnetic BEC in the absence of 
a confining trap. Now, $n_1'=n_{-1}'=1/(2L)$ are the maximum density values 
allowed for these two components of the system with zero magnetization 
for the immiscible case. Any general distribution with zero 
magnetization for the immiscible profile will have, due to the inherent 
symmetry of the present model between components $j=1$ and -1, $n_1' = 1/(2L)-\delta$ 
between $x=-L$ to $0$, $n_{-1}' = 1/(2L)-\delta$ between $x=0$ to $L$, and 
$n_0'=\delta$ between $x=-L$ to $L$, with $\delta\ge 0$. The interaction 
energy corresponding to this general distribution for the immiscible profile
is
\begin{equation}
 E_{\rm int} = N\left[\frac{c_0-|c_2|}{4L}+|c_2|\delta^2L\right].
\end{equation}
Hence, the interaction energy for this immiscible profile is either more than 
$(\delta>0)$ or equal to $(\delta=0)$ the interaction energy of the  miscible
one. Hence for a general distribution ($\delta\ne 0$) the miscible profile with 
the lowest energy will be the preferred ground state. The presence of a 
trapping potential, however small it may be, will favor the miscible profile 
due to an extra confining force to the center.

Now let us consider the phase separation in a polar BEC. Interaction energy 
(\ref{xxx}) can be minimized if we choose 
\begin{equation}
n_1 = n_{-1} = \frac{1}{4L}, n_{0}= 0
\label{den_mis_polar1}
\end{equation} 
in the case of a miscible profile [viz. Eqs. (\ref{ma})-(\ref{mc})] or 
$n_1=n_{-1} = 0, n_0 = 1/(2L)$ in the case of an immiscible profile 
[viz. Eqs. (\ref{xa})-(\ref{xc})]. This immiscible profile represents 
effectively a single component system. The value of the minimum energy 
in both the cases is 
\begin{equation}
 E_{\rm int}^{\rm min} = \frac{Nc_0}{4L}.
\end{equation}
As mentioned in Sec. \ref{Sec-II}, the phase-separation in polar 
condensate is most likely to occur between the  $0$th and
overlapping $1$ and $-1$ components.
Therefore, we also consider the profile where the components $1$ and $-1$ are 
miscible, and these two are phase separated from the $0$th component with
the following general density distribution 
\begin{eqnarray}
  \rho_1(x)  &=& \left\{
     \begin{array}{lr}
        n_1'', &  \quad  \quad\quad \quad  -L<x<-L+L',\\
        0, & \quad \quad \quad \quad  ~-L+L'\leq x\leq L,
     \end{array}
   \right.\label{ps_polar1}\\
   \rho_0(x)  &=& \left\{
     \begin{array}{lr}
        0, & \quad \quad \quad \quad  ~-L < x< -L+L',\\
        n_0'', & \quad \quad \quad \quad  \quad   -L+L'\leq x\leq L,
     \end{array}
   \right.\label{ps_polar2}
\\
   \rho_{-1}(x)  &=& \left\{
     \begin{array}{lr}
        n_{-1}''=n_1'', &  ~-L<x<-L+L',\\
        0, &  ~-L+L'\leq x\leq L,
     \end{array}
   \right.\label{ps_polar3}
\end{eqnarray}
where $L'<2L$, and all the densities are zero for $|x|>L$.
The interaction energy for this
distribution is
\begin{eqnarray}
 E_{\rm int} = \frac{Nc_0}{2}[4(n_1'')^2L'+2(n_0'')^2L-(n_0'')^2L'],
 \end{eqnarray}
with  the normalization condition
 \begin{equation}
  2n_1''L'+2n_0L-n_0L' =1.
  \label{norm3}
 \end{equation}
The minimum of this energy, subject to the normalization constraint, occurs at
\begin{equation}
L' = L,~n_1'' = n_{-1}'' = \frac{1}{4L},~\rm{and}~n_0'' = \frac{1}{2L}. 
\end{equation} The minimum interaction energy for this density distribution 
is the same as for the miscible profile, i.e., $E_{\rm int}^{\rm min} =Nc_0/(4L)$. 
Similarly, it can be shown that the profile where all the three components 
are phase separated from each other as well as the rest of the possible 
phase separated profiles, the interaction energy is always greater than 
$Nc_0/(4L)$ due to a non-zero contribution from the $c_2$ dependent terms. 
So, the energy of any general immiscible profile is either equal to or greater 
than $Nc_0/(4L)$ due to a non-zero contribution from the $c_2$ dependent terms.
The presence of a trapping potential, however weak it may be, will make
the miscible profile energetically more favorable than the all possible 
immiscible profiles. Hence, there can be no phase separation in the trapped 
ferromagnetic and polar BECs. 

Next let us consider the effect of the SO coupling and the Rabi term on a phase 
separation. First, let us include the SO coupling without the Rabi term 
$(\gamma\ne0,\Omega=0)$ and  discuss the effect on a ferromagnetic BEC 
$(c_2<0)$. The presence of this term leads to a constant phase gradient 
$-\alpha$ and $\alpha$ in $\phi_1$ and $\phi_{-1}$, respectively \cite{Bao}. 
The interaction energy of the  miscible profile 
[viz. Eqs. (\ref{ma})-(\ref{mc})] in this case is
\begin{eqnarray}
 E_{\mathrm{int}} &=& NL\left[c_0(4n_1^2+n_0^2+4n_1n_0) -|c_2|8n_1n_0\right.\nonumber\\
   &&\left.+ 2\alpha^2n_1-4\gamma\alpha n_1 \right],
\end{eqnarray}
where the  $  2N\alpha^2n_1L$ term arises from the derivatives of the phases of 
$\phi_1$ and $\phi_{-1}$. Minimizing this energy with respect to $n_1$ and 
$\alpha$, subject to the normalization constraint  ~(\ref{norm1}) and 
$n_j\ge0$ for $j=1,0,-1$, we get
\begin{equation}\label{n1}
 \alpha = \gamma,~
  n_1 = \left\{
     \begin{array}{lr}
        \frac{1}{8L} +\frac{\gamma^2}{16|c_2|}, 
         &   \gamma\le\sqrt{2 |c_2|/L}\\
         \frac{1}{4L},  &  \gamma>\sqrt{2 |c_2|/L},
     \end{array}
   \right.
\end{equation}
with the corresponding minimum energy 
\begin{align} 
   E^{\rm min(M)}_{\mathrm{int}} = \left\{
     \begin{array}{lr}
       N\left[\frac{c_0-|c_2|}{4L} -\frac{\gamma^2}{4}-\frac{L\gamma^4}{16|c_2|}\right], 
         &  \gamma\le\sqrt{2 |c_2|/L}\\
       N\left[\frac{c_0}{4l}-\frac{\gamma^2}{2}\right],
         &  \gamma>\sqrt{2 |c_2|/L}.\label{E_m_ferro}
     \end{array}
   \right.
 \end{align} 
The density $n_1$ of Eq.~(\ref{n1}) attains a saturation for 
$\gamma>\sqrt{2|c_2|/L}$. With further increase of $\gamma$ the 
density $n_1$ does not change as it has already achieved the maximum
permissible density for a state with ${\cal M}=0$ subject to the
normalization constraint ~(\ref{norm1}).

The interaction energy of the  immiscible profile 
[viz. Eqs. (\ref{xa})-(\ref{xc})] in this case is
\begin{eqnarray}
 E_{\mathrm{int}} &=&  NL[c_0(n_1'^2+n_0'^2+2n_1'n_0')
                       -|c_2|n_1'(n_1'+2n_0')\nonumber\\
    &&+ \alpha'^2n_1'-2\gamma\alpha' n_1' ],
\end{eqnarray}
Minimizing this energy with respect to $n_1'$ and $\alpha'$, subject to the 
normalization constraint ~(\ref{norm2}) and $n_j\ge0$ for $j=1,0,-1$, we get
\begin{equation}
 \alpha' = \gamma,~n_1' = \frac{1}{2L},
 \label{den_imis_ferro}
\end{equation}
with the corresponding  minimum energy
\begin{equation}
 E^{\rm min(I)}_{\mathrm{int}} = N\left[\frac{c_0-|c_2|}{4L} -\frac{\gamma^2}{2}\right].
 \label{E_im_ferro}
\end{equation}
Comparing Eqs.~(\ref{E_m_ferro}) and (\ref{E_im_ferro}), we find that the 
immiscible profile has lower energy than the miscible one for any non-zero
value of $\gamma$ for a ferromagnetic BEC: 
$E_{\rm{int}(\Omega)}^{\rm min(I)}<E_{\rm{int}(\Omega)}^{\rm min(M)}$. 
Hence the SO coupling will favor phase separation in a  ferromagnetic 
BEC.

Let us now discuss the phase separation in a polar BEC in the presence of 
a SO coupling. The interaction energy of the  miscible profile 
[viz. Eqs. (\ref{ma})-(\ref{mc})] in this case is
\begin{eqnarray}
 E _{\rm{int}}&=& NL[c_0(4n_1^2+n_0^2+4n_1n_0)+8c_2n_1n_0 \nonumber\\
   &&+ 2\alpha^2n_1-4\gamma\alpha n_1] .
\end{eqnarray}
Minimizing it, subject to the normalization constraint ~(\ref{norm1}) and 
$n_j\ge0$, we get
\begin{equation}
 \alpha = \gamma,~n_1 = n_{-1} = \frac{1}{4L},~ n_0 = 0. 
 \label{den_mis_polar2}
\end{equation}
The value of the  minimum energy for this miscible profile is
\begin{equation}
 E^{\rm min(M)} _{\rm{int}} = N\left[\frac{c_0}{4L} -\frac{\gamma^2}{2}\right].
 \label{E_m_polar}
\end{equation}
Similarly, the energy of the immiscible profile 
[viz. Eqs. (\ref{xa})-(\ref{xc})] of the polar BEC is
\begin{eqnarray}
 E _{\rm{int}}&=& NL[c_0(n_1'^2+n_0'^2+2n_1'n_0')+c_2n_1'(n_1'+2n_0')
          \nonumber\\
      && + \alpha'^2n_1'-2\gamma\alpha' n_1' ],
\end{eqnarray}
Minimizing this energy, subject to the normalization constraint ~(\ref{norm2}) 
and $n_j\ge0$, we get 
\begin{equation}
   n_1' = n_{-1}'=\left\{
     \begin{array}{lr}
       \frac{1}{2L}, &   \gamma>\sqrt{c_2/(2L)}\\
        0, &   \gamma\le\sqrt{c_2/(2L)},
     \end{array}
   \right.
\end{equation}
with the corresponding minimum energy given by 
\begin{equation}
   E^{\rm min(I)}  _{\rm{int}}
= \left\{
     \begin{array}{lr}
       N\left[\frac{c_0+c_2}{4L}-\frac{\gamma^2}{2}\right], &  \gamma>\sqrt{c_2/(2L)}\\
       \frac{Nc_0}{4L}, &  \gamma\le\sqrt{c_2/(2L)}.
     \end{array}
   \right.
\end{equation}
This energy is larger than the energy of the miscible profile given by 
Eq. ~(\ref{E_m_polar}): 
$E^{\rm min(I)}  _{\rm{int}}>  E^{\rm min(M)}  _{\rm{int}}$.
Similarly, it can be argued that the energies of the other possible immiscible 
profiles with $n_0\ne0$, like the distribution in 
Eqs. (\ref{ps_polar1})-(\ref{ps_polar3}), are always larger than 
$N(c_0-2\gamma^2L)/(4L)$ due to an increase in the negative energy contribution 
from the $\gamma$ dependent term, i.e., this contribution is larger than 
$-\gamma^2/2$. Hence, the SO coupling will favor miscibility in the case of a 
polar BEC.

Now let us analyze the role of the Rabi term ($\Omega \ne 0$). For the sake 
of simplicity let us assume that $\gamma = 0$. The energy contribution from 
the Rabi term is
\begin{eqnarray}
   E_{\rm{int}(\Omega)} &=&  \sqrt{2\rho_0(x)}\Omega N\int_{-\infty}^{\infty}
                            \left[ \sqrt{\rho_1(x)}\cos(\theta_0-\theta_1)\right.
                            \nonumber\\
             &&\left.+\sqrt{\rho_{-1}(x)}\cos(\theta_0-\theta_{-1})\right]dx.
\end{eqnarray}
This expression is valid in general for nonuniform densities and not just 
in the case of  uniform densities appropriate for  the 1D box.
This term will lead to a decrease in   energy of the system
if 
\begin{eqnarray}
 \frac{\pi}{2}<|\theta_0-\theta_1|<\frac{3\pi}{2}, \quad \mbox{and} \quad 
 \frac{\pi}{2}<|\theta_0-\theta_{-1}|<\frac{3\pi}{2}.
\end{eqnarray}
Assuming that $\theta_0 = 0$, the minimum of $E_{\rm{int}(\Omega)}$ for 
the miscible profile [viz. Eqs. (\ref{ma})-(\ref{mc})] occurs at 
\begin{equation}
 n_1 = n_{-1} = \frac{1}{8L},\quad  n_0 = \frac{1}{4L},
                \quad |\theta_1| = |\theta_{-1}| = \pi,
\end{equation}
The value of the corresponding minimum energy is  
$E_{\rm{int}(\Omega)}^{\rm min(M)} = -N\Omega$. The minimum for the immiscible 
profile [viz. Eqs. (\ref{xa})-(\ref{xc})] occurs at 
\begin{equation}
   n_1 = n_{-1} = \frac{1}{4L},\quad n_0 = \frac{1}{4L},
                  \quad|\theta_1| = |\theta_{-1}| = \pi,
\end{equation}
with the corresponding energy minimum $E_{\rm{int}(\Omega)}^{\rm min(I)}=  
-N\Omega/\sqrt{2}$. 
Also, the $E_{\rm{int}(\Omega)}$ of the distribution represented by 
Eqs.~(\ref{ps_polar1})-(\ref{ps_polar3}) is uniformly zero and hence greater 
than $-N\Omega$. 
Hence, the Rabi term favors miscibility in the spinor BEC irrespective of the 
nature of spin interaction: ferromagnetic or polar. It implies that in a 
ferromagnetic BEC the terms containing $\gamma$ (favoring phase separation) 
and $\Omega$ (favoring miscibility) will have opposite  roles as far as 
phase separation is concerned.


\section{Spinor BEC in a harmonic trap}
\label{Sec-IV}

In the presence of a harmonic trap, we study the ground state structure of the 
spinor BEC by solving Eqs.~(\ref{gp_s1}) - (\ref{gp_s3}) numerically. We use 
split-time-step finite-difference method to solve the coupled 
Eqs.~(\ref{gp_s1}) - (\ref{gp_s3}) \cite{Muruganandam,H_Wang}. The spatial and 
time steps employed in the present work are $\delta x = 0.05$ and 
$\delta t = 0.000125$. In order to find the ground state, we solve 
Eqs.~(\ref{gp_s1}) - (\ref{gp_s3}) by imaginary-time propagation. 
{The imaginary
time propagation neither conserves norm nor magnetization.} 
To fix both norm and magnetization, we use the method elaborated in 
Ref.~\cite{Bao}. Accordingly, after each iteration in imaginary time 
$\tau = -it$, the wave-function components are transformed as
\begin{equation}
 \phi_j(x,\tau+d\tau) = d_j \phi_j
(x,\tau),
\end{equation}
where $d_j$'s with $j=1,0,-1$ are the normalization constants. Now, the 
chemical potential of the three components are related as
\begin{equation}
\mu_1+\mu_{-1} = 2\mu_0.
\end{equation}
Using this relation, one can derive the relation between the three normalization 
constants \cite{Bao}:
\begin{equation}
d_1d_{-1} = d_0^2.
\label{liv}
\end{equation}
Using Eq.~(\ref{liv}) along with the normalization [viz. Eq.~(\ref{norm})] and 
magnetization constraints [viz. Eq.~(\ref{magnetization})], $d_j$'s can be 
determined as \cite{Bao}
\begin{align}
 d_0 &=\frac{\sqrt{1-{\cal M}^2}}{\sqrt{N_0+\sqrt{4(1-{\cal M}^2)
       N_1N_{-1}+{\cal M}^2N_0^2}}},\\
 d_1 &=\sqrt{\frac{1+{\cal M}-c_0^2N_0}{2N_1}},\\
 d_{-1} &=\sqrt{\frac{1-{\cal M}-c_0^2N_0}{2N_{-1}}},
\end{align}
and here $N_j = \int |\phi_j(x,\tau)|^2 dx$. These normalization constants
ensure that the norm and magnetization are both conserved after each iteration
in imaginary time.
The quasi-1D trap considered here  has $\omega_x = 2\pi\times20$ Hz, 
$\omega_y =\omega_z = 2\pi\times400$ Hz. We consider $^{87}$Rb atoms 
with $a_0=5.387$ nm and $a_2 = 5.313$ nm as a typical example of ferromagnetic 
BEC. As a polar BEC, we consider $^{23}$Na which has $a_0=2.646$ nm and 
$a_2 = 2.911$ nm. The values of $l_0$  are $2.41~\mu$m and $4.69~\mu$m 
for $^{87}$Rb and $^{23}$Na, respectively.

\begin{figure}[!t]
\begin{center}
\includegraphics[trim = 1cm 0mm 0cm 0mm, clip,width=.8\linewidth,clip]{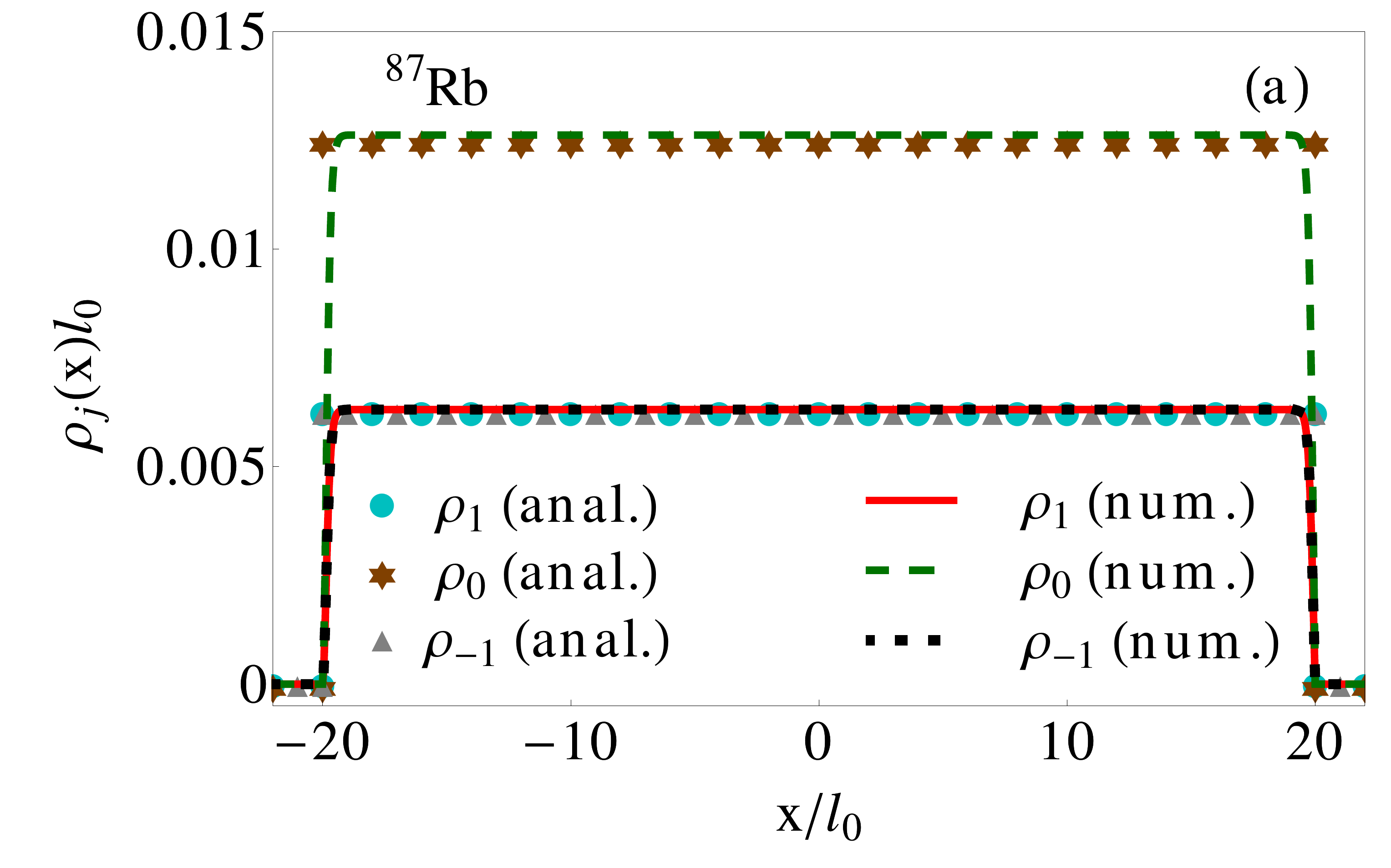}\\
\includegraphics[trim = 0mm 0mm 0cm 0mm, clip,width=.84\linewidth,clip]{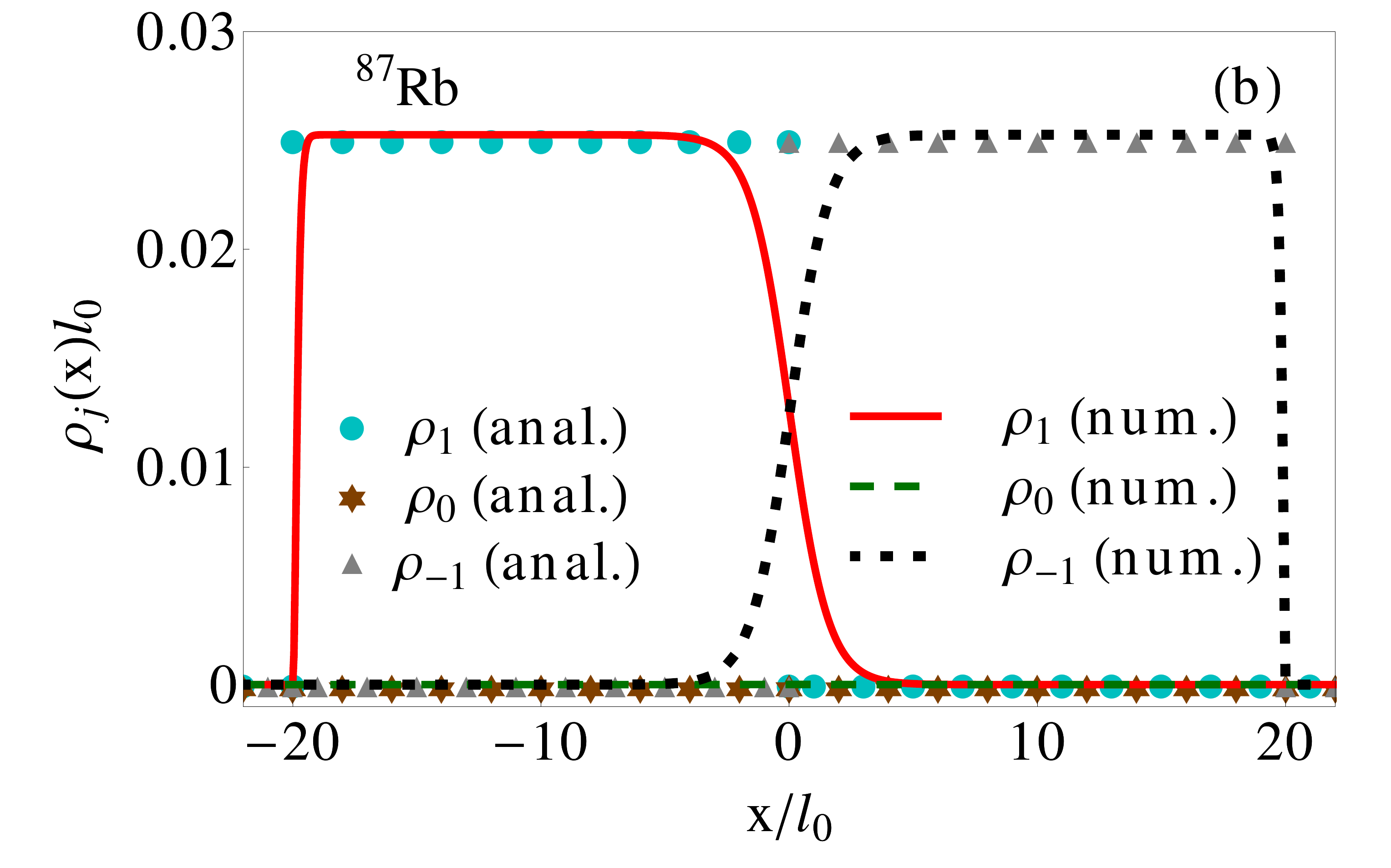}\\
\includegraphics[trim = 1cm 0mm 0cm 0mm, clip,width=.8\linewidth,clip]{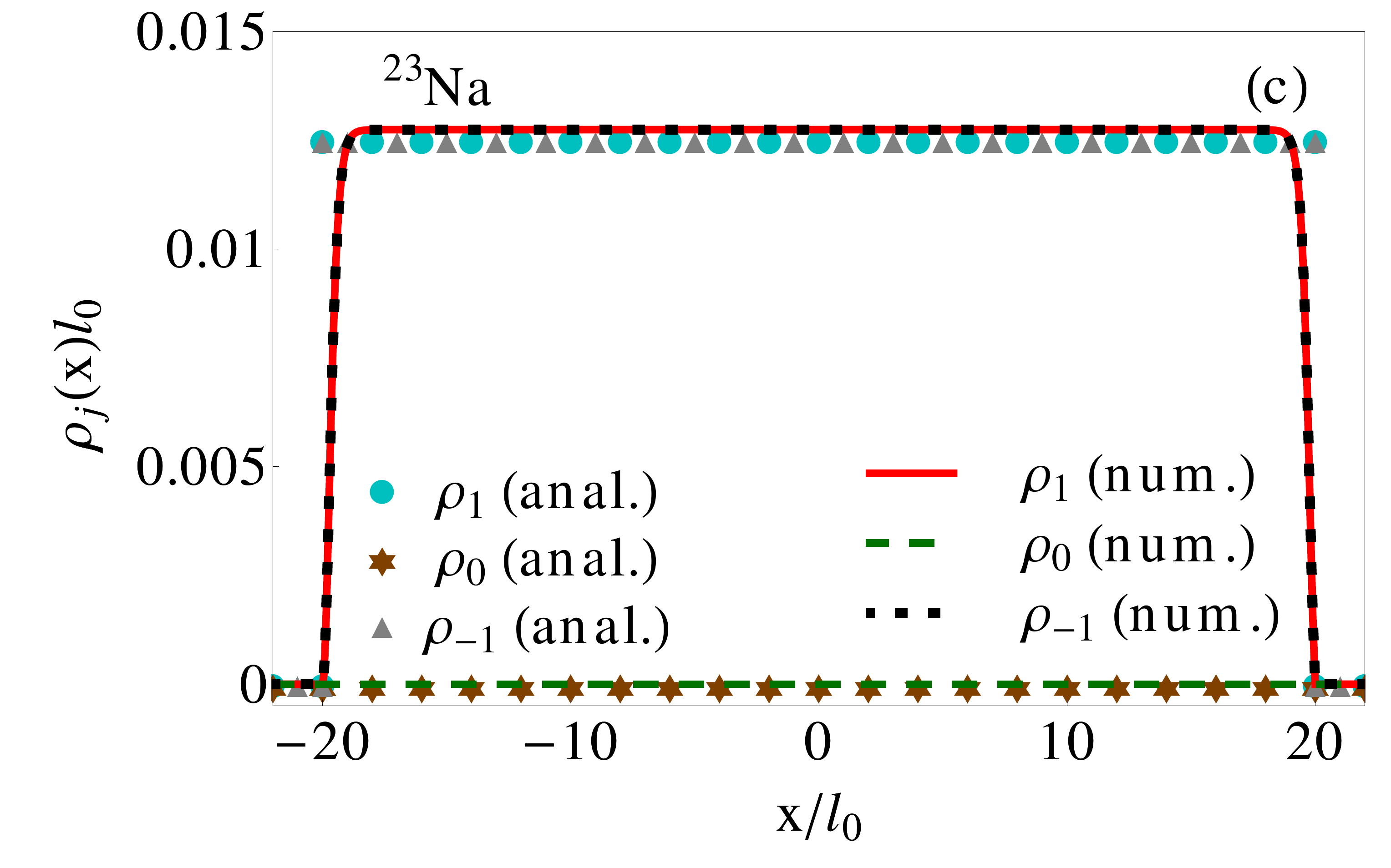}
\caption{(Color online) (a) and (b) Analytical (anal.) and numerical (num.) densities $\rho_j(x)l_0$
of a condensate of $^{87}$Rb atoms with $c_0 = 885.72l_0$ and $c_2=-4.09l_0$ in a 1D
box of length $40l_0$. The SO coupling $\gamma=0$ and $0.5$ for (a) and (b), 
respectively. (c) The same for  a condensate of $^{23}$Na atoms with $c_0 = 241.28l_0$ 
and $c_2=7.76l_0$ in the presence of an arbitrary SO coupling. Both the densities and 
spatial coordinates in this figure are in dimensionless units.}
\label{box_pot_res} \end{center}
\end{figure}

Before proceeding to the numerical solutions of the spinor condensate in
a harmonic trap, let us first compare the analytic results for
the condensate in a 1D box with the corresponding numerical ones.
For this purpose, we consider aforementioned oscillator lengths for $^{87}$Rb
and $^{23}$Na in a 1D box of length $40l_0$. The non-linearities 
$(c_0,~c_2)$  considered for $^{87}$Rb and $^{23}$Na are, respectively, 
$(885.72l_0,-4.09l_0)$ and $(241.28l_0,7.76l_0)$.  In Fig.~\ref{box_pot_res} (a), analytic and numerical 
densities
for the $^{87}$Rb condensate  in the absence of SO coupling and Rabi term, given by 
Eq.~(\ref{den_mis_ferro}), have been plotted.
In Fig.~\ref{box_pot_res} (b), analytic and numerical densities for the $^{87}$Rb 
condensate in the presence of SO coupling ($\gamma = 0.5,\Omega = 0$), 
given by Eq.~(\ref{den_imis_ferro}), are shown. Finally,  in Fig.~\ref{box_pot_res} (c),  the same  for 
the $^{23}$Na condensate in the absence as well as presence an arbitrary SO 
coupling, given by Eqs.~(\ref{den_mis_polar1}) and (\ref{den_mis_polar2}), have been illustrated.
We find that the numerical
results are in good agreement with the analytic predictions as is shown in
Fig.~\ref{box_pot_res}.

\begin{figure}[!t]
\begin{center}
\includegraphics[trim = 3mm 0mm 1cm 0mm, clip,width=.49\linewidth,clip]{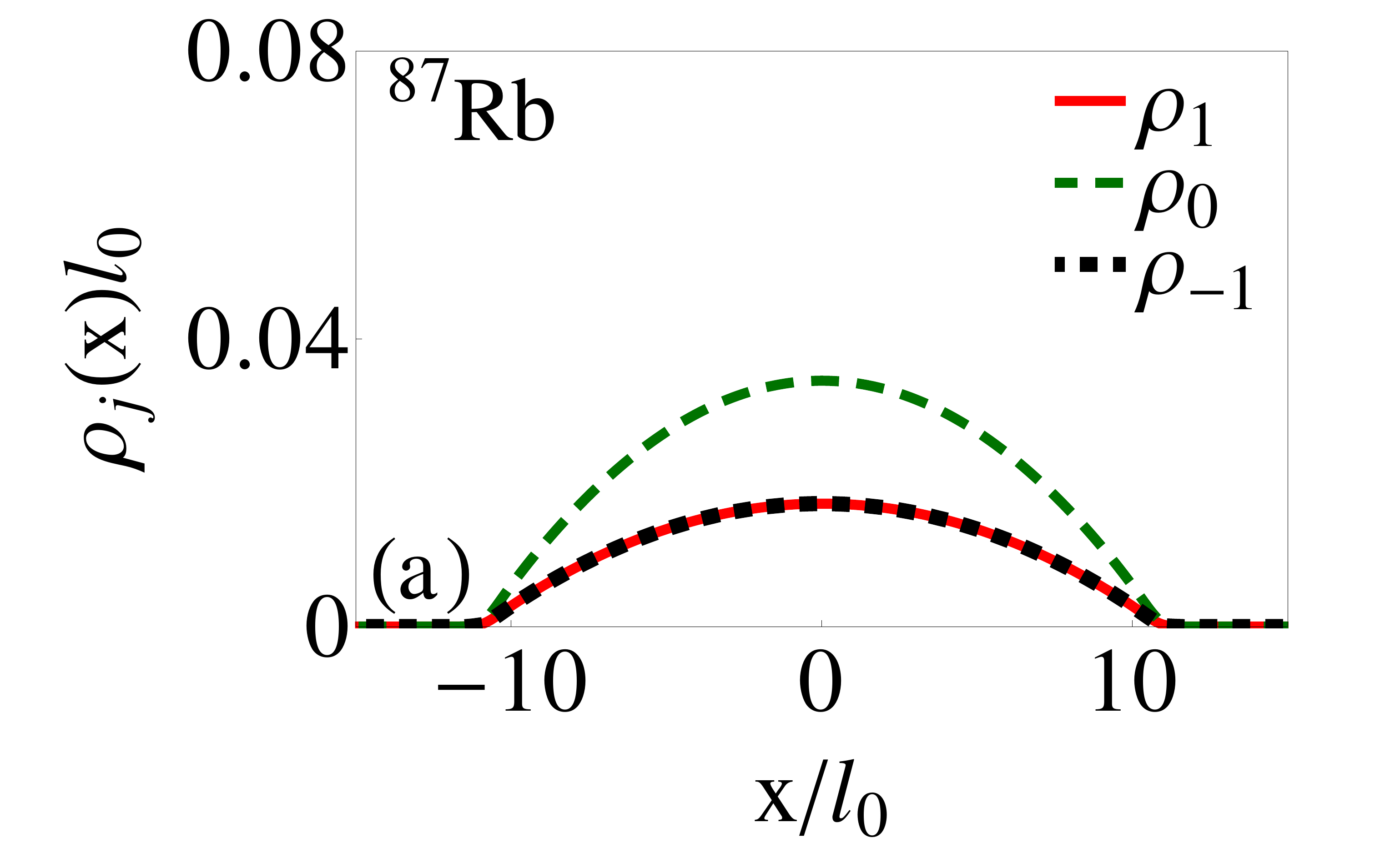}
\includegraphics[trim = 3mm 0mm 1cm 0mm, clip,width=.49\linewidth,clip]{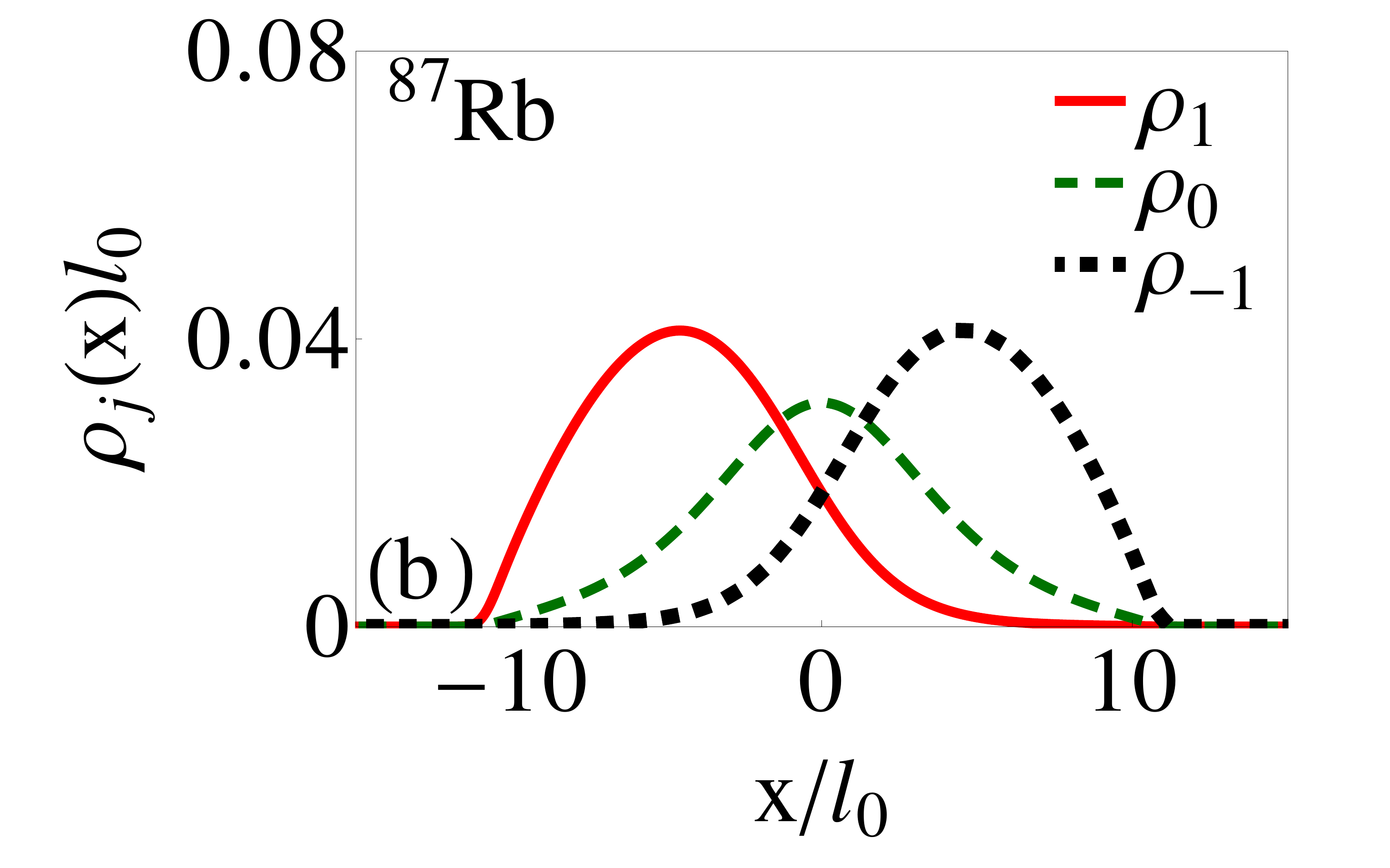}
\includegraphics[trim = 3mm 0mm 1cm 0mm, clip,width=.49\linewidth,clip]{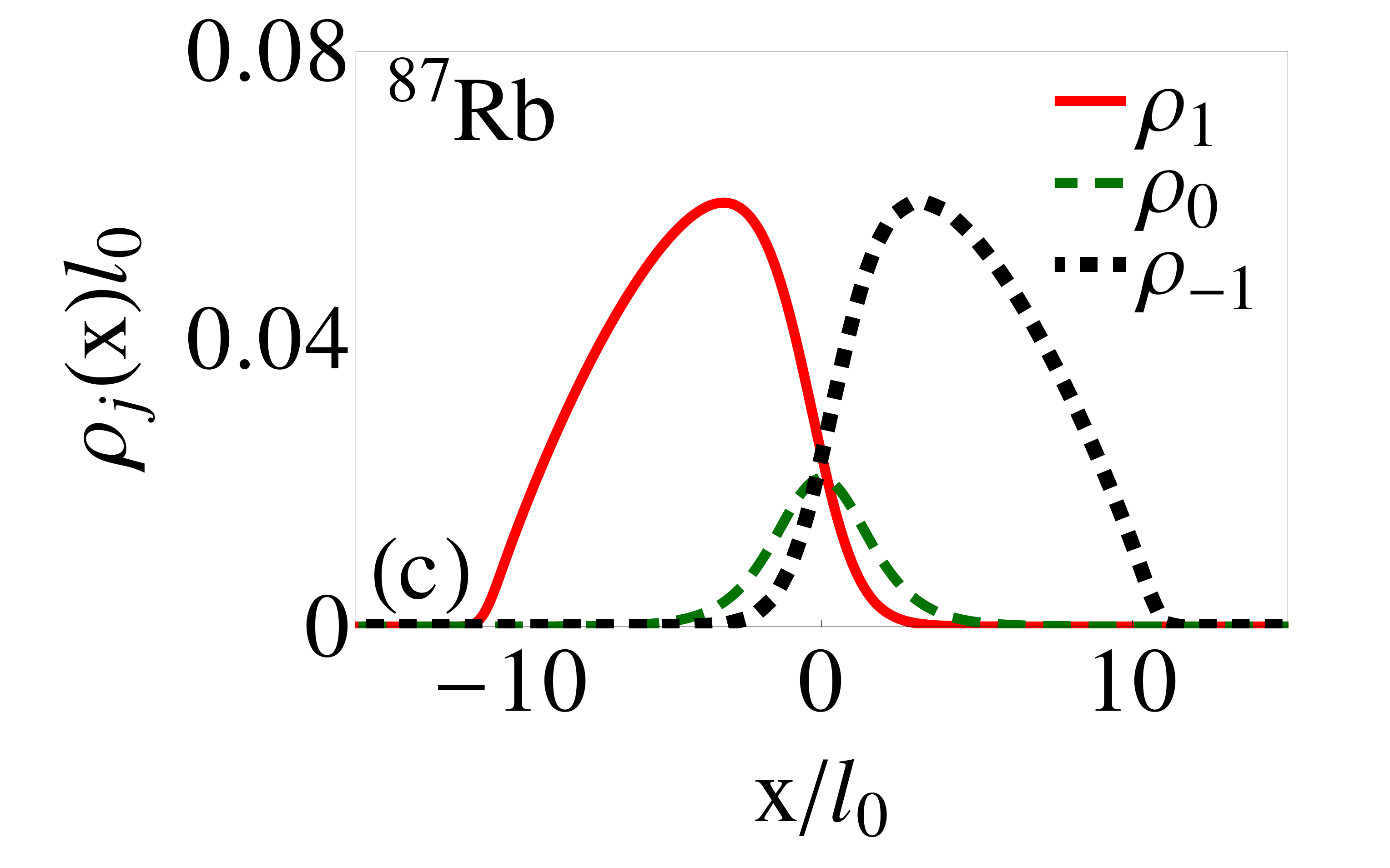}
\includegraphics[trim = 3mm 0mm 1cm 0mm, clip,width=.49\linewidth,clip]{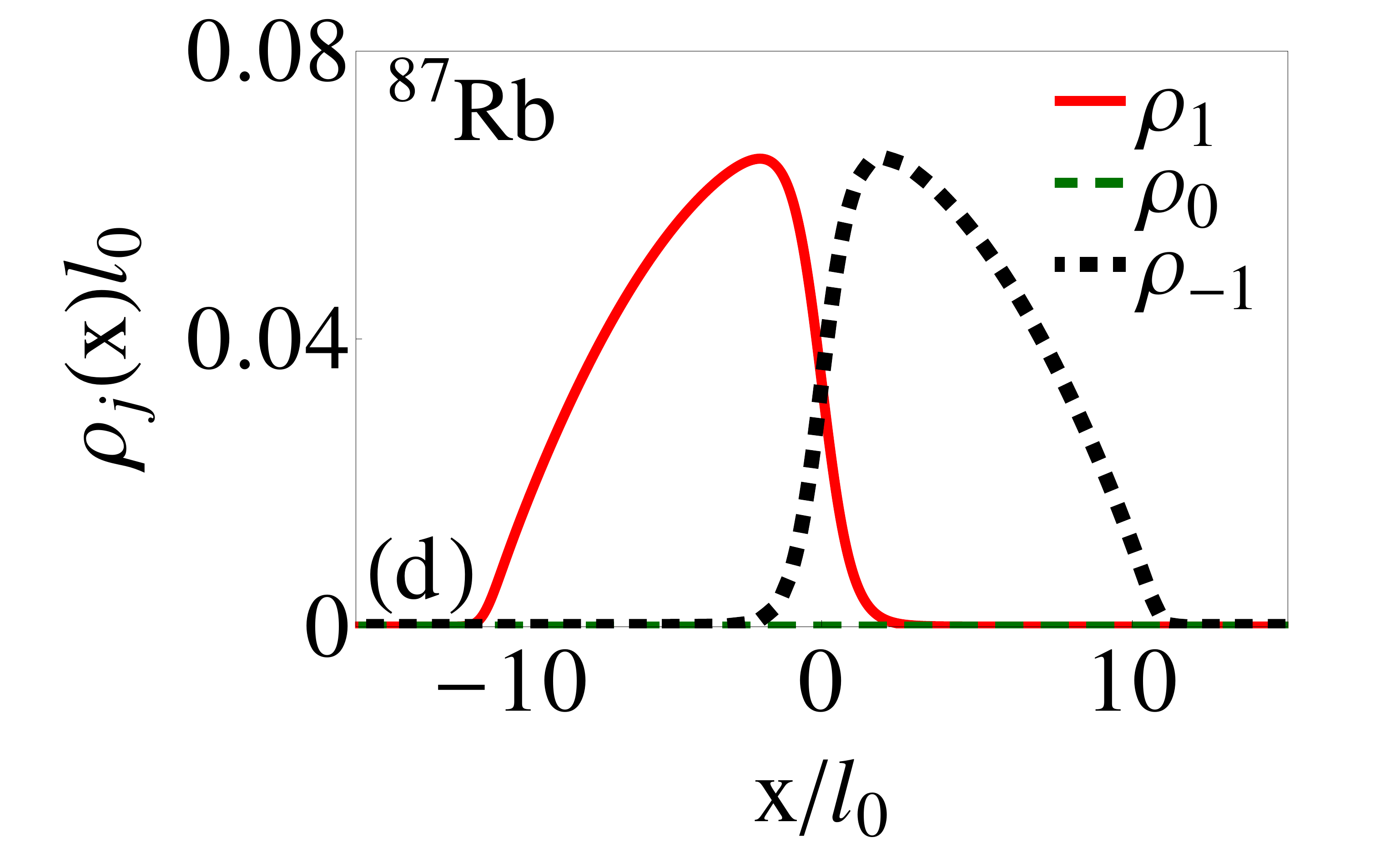}
\caption{(Color online) Ground state structure of $^{87}$Rb spinor BEC with 
$10000$ atoms with $\Omega=0$. The SO coupling 
$\gamma = 0,~0.25,~0.5,~1$ for (a), (b), (c) and (d) respectively. All 
quantities in this figure are dimensionless.}
\label{fig-1} \end{center}
\end{figure}

\begin{figure}[!b]
\begin{center}
\includegraphics[trim = 3mm 0mm 1cm 0mm, clip,width=.49\linewidth]{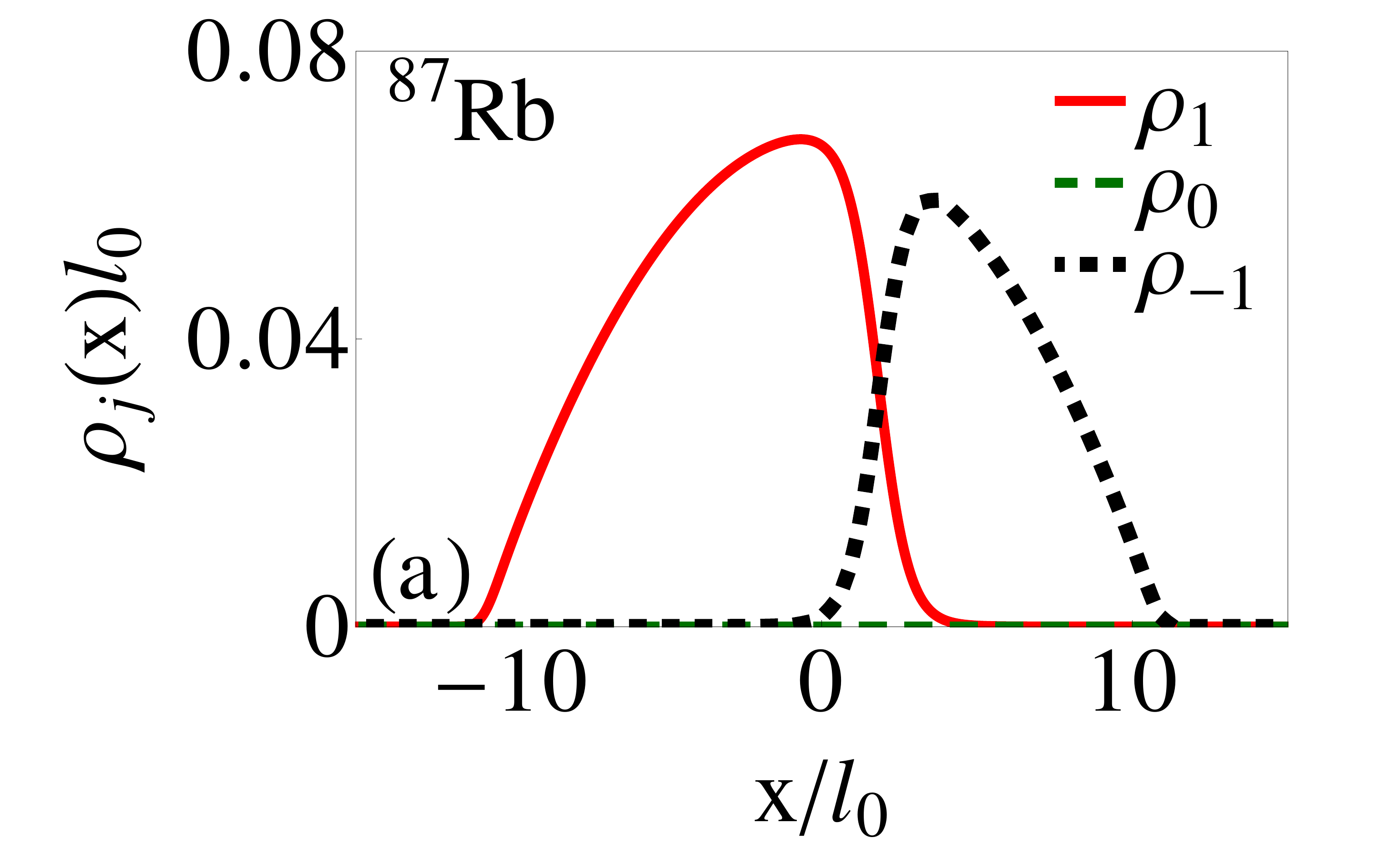}
\includegraphics[trim = 3mm 0mm 1cm 0mm, clip,width=.49\linewidth]{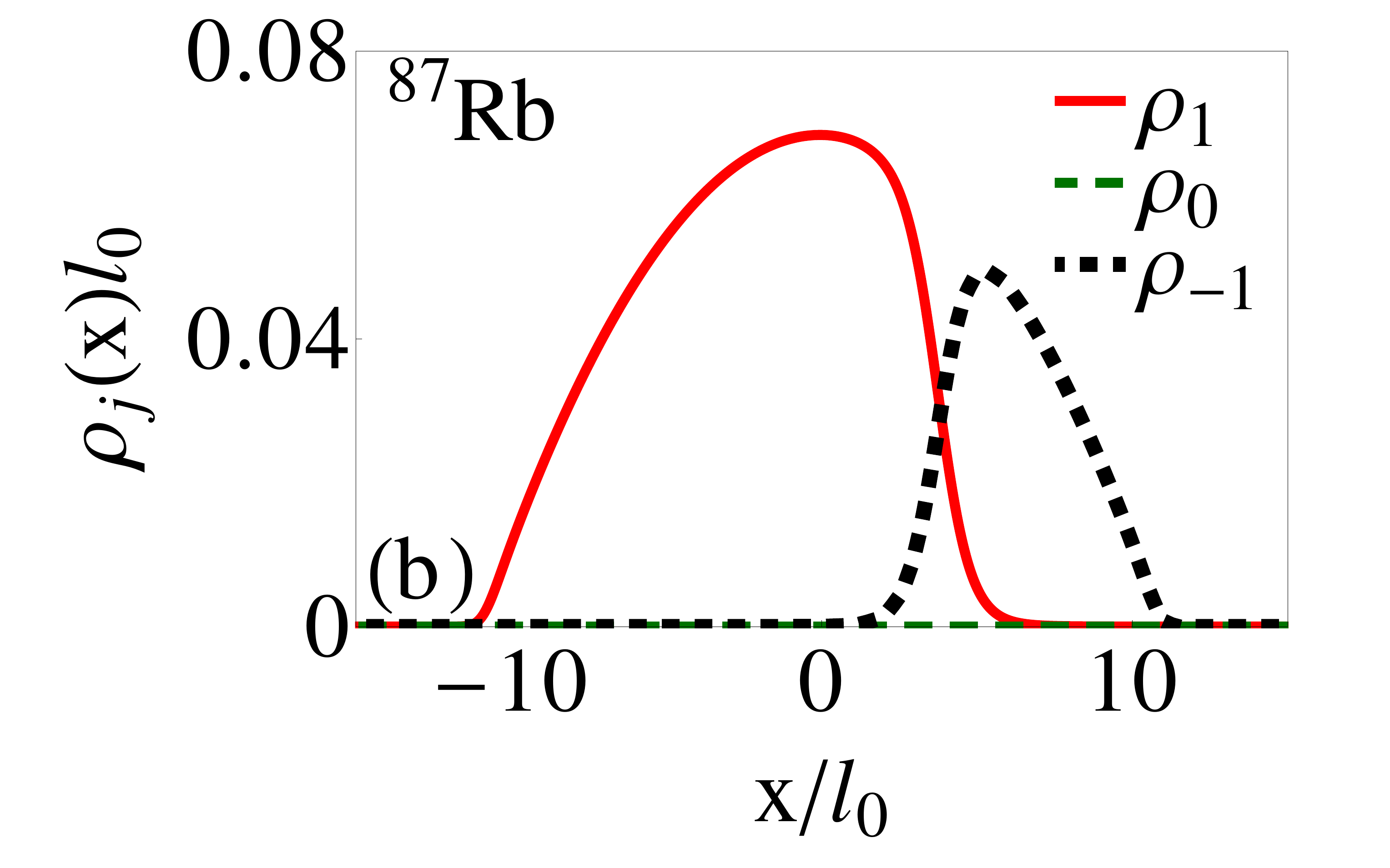}
\includegraphics[trim = 3mm 0mm 1cm 0mm, clip,width=.49\linewidth]{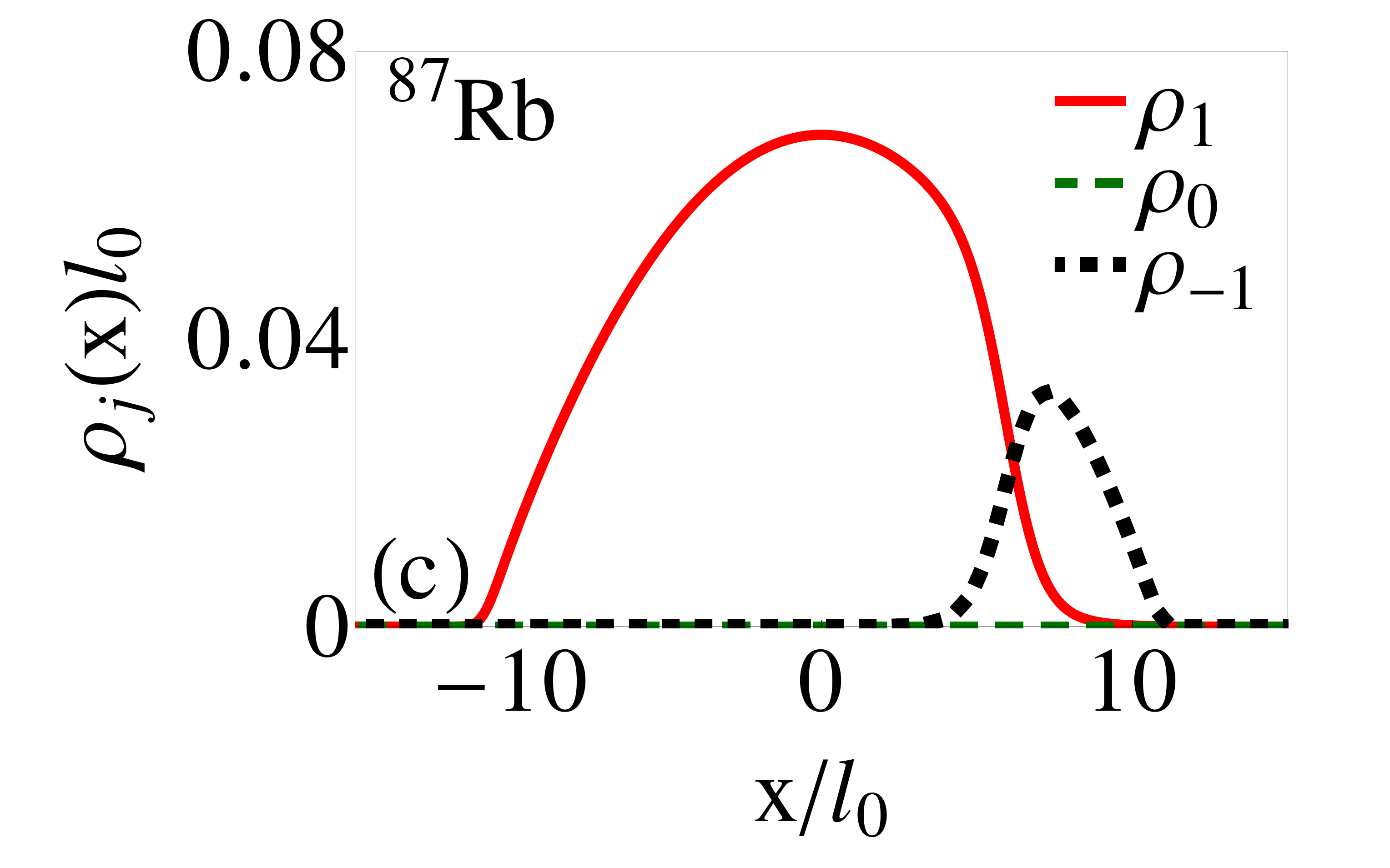}
\includegraphics[trim = 3mm 0mm 1cm 0mm, clip,width=.49\linewidth]{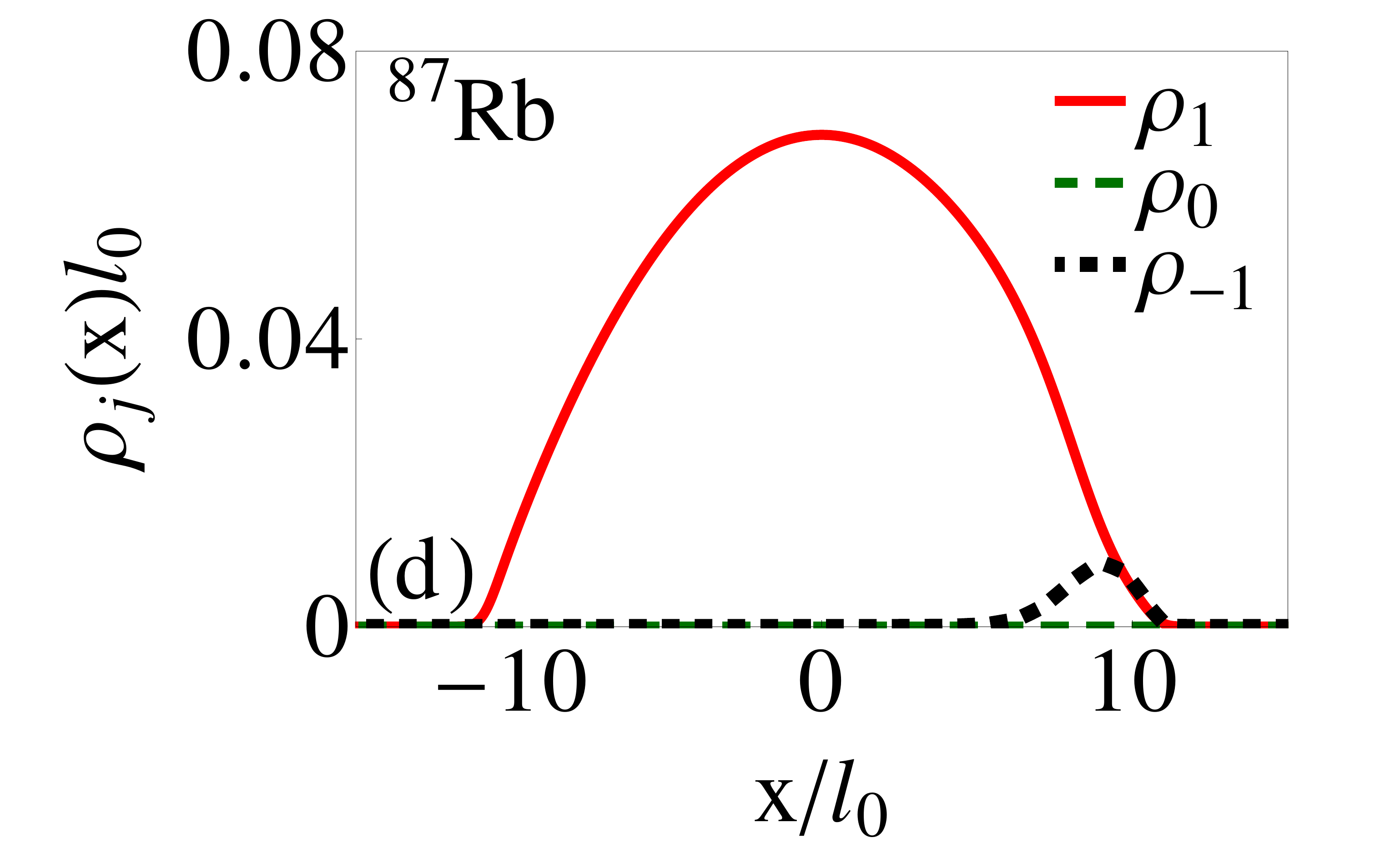}

\caption{(Color online) Ground state structure of $^{87}$Rb spinor BEC with 
$10000$ atoms with $\Omega=0$, $\gamma = 1$. Both the density and spatial coordinates are
plotted in dimensionless units. The magnetization ${\cal M}
 =0.25$, $0.5$, $0.75$, $0.95$ for (a), (b), (c) and (d) 
respectively. 
}
\label{fig-2} \end{center}
\end{figure}

 Now, let us discuss the harmonically trapped spinor condensates.
In Fig. \ref{fig-1} we present the densities of the ground state of 
$10000$ $^{87}$Rb atoms with different   SO coupling and 
without the Rabi term. Without the SO coupling, the ground state solution 
for $^{87}$Rb is miscible and $\rho_0(x)>\rho_{1}(x)=\rho_{-1}(x)$ for zero 
magnetization (${\cal M} = 0$) [viz. Fig.~\ref{fig-1}(a)], which is in 
qualitative agreement with the conclusion of the analytic study of the 
uniform system in Sec. \ref{Sec-III} given by  Eq. (\ref{den_mis_ferro}). 
If the number of atoms is sufficiently large, as the   the SO 
coupling $\gamma$ is increased, the density  $\rho_0$ starts decreasing 
slowly, which ultimately makes the system immiscible as is shown in
Figs.~\ref{fig-1}(a) - (d). For a sufficiently strong SO coupling, $\rho_0$ 
becomes zero and there is a maximum of phase separation between the two 
remaining component densities $\rho_1(x)$ and $\rho_{-1}(x)$. This is again 
in agreement with the result of the analytic study on the uniform system 
given by Eq.~(\ref{den_imis_ferro}), which predicts zero density for the 
$0$th component. However, if the number of atoms is smaller 
($N\le 1000$), the $0$th component again vanishes with the increase in SO 
coupling $\gamma$ above a critical value, but there is no phase separation 
between components $1$ and $-1$.

{ The state with $\rho_0(x)=0$ appears naturally with the increase of the
SO coupling, and this in a zero magnetization  case guarantees 
an equal number of atoms for the components $1$ and $-1$ resulting in  
$\rho_1(x) =\rho_{-1}(-x)$.} It is interesting to study the fate of this 
state as the magnetization is increased (${\cal M} > 0 $). Keeping 
$\gamma = 1$ and $\Omega = 0$ fixed, one can change the relative proportion 
of   $\rho_1$ and $\rho_{-1}$ by changing the magnetization ${\cal M}$, as 
is shown in Fig.~\ref{fig-2} (a) - (d), maintaining $\rho_0 (x)=0$.  
With increasing ${\cal M}$ the relative density of component $-1$ decreases 
and the system turns miscible from immiscible.

\begin{figure}[!t]
\begin{center}
\includegraphics[trim = 3mm 0mm 1cm 0mm, clip,width=.49\linewidth]{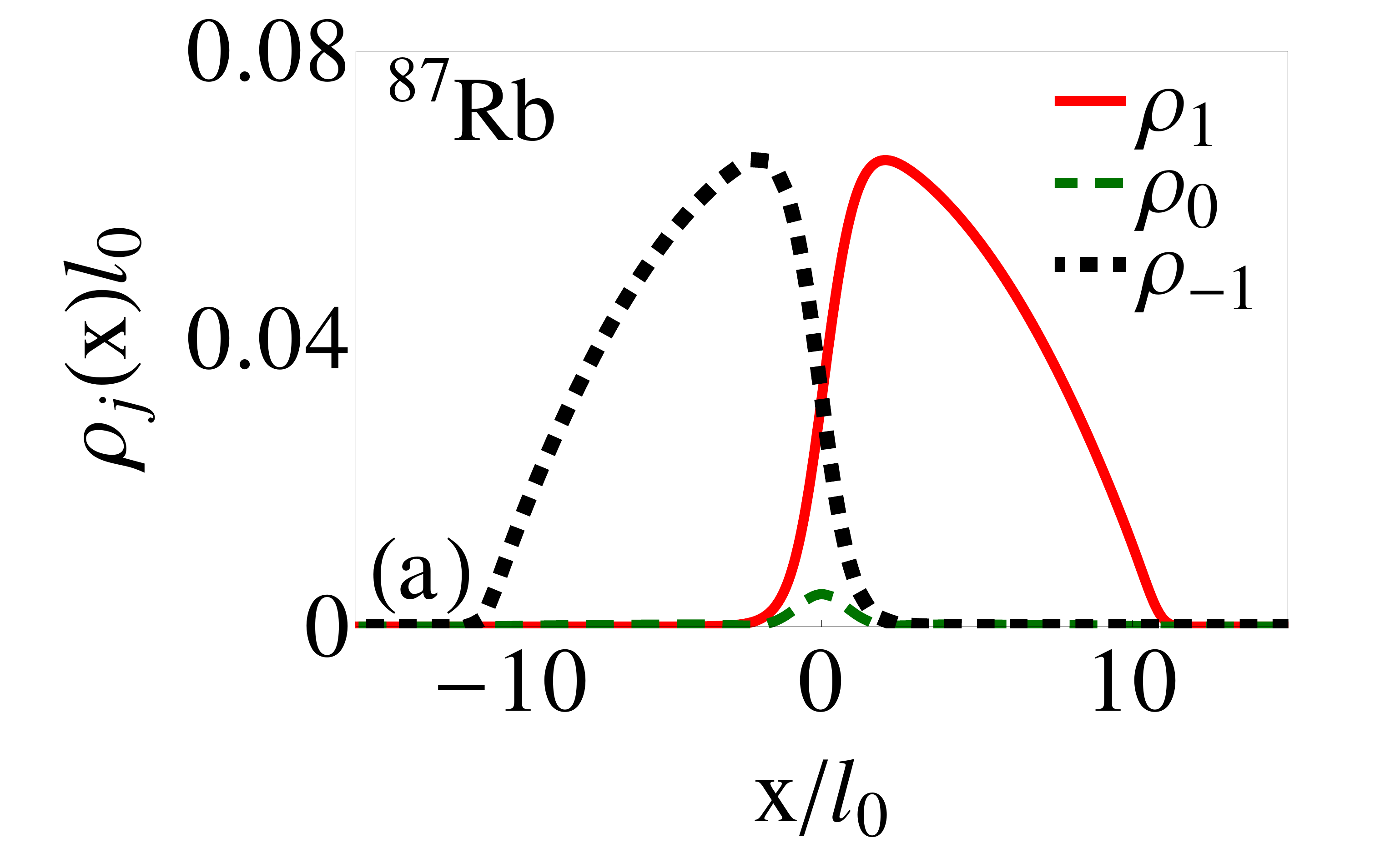}
\includegraphics[trim = 3mm 0mm 1cm 0mm, clip,width=.49\linewidth]{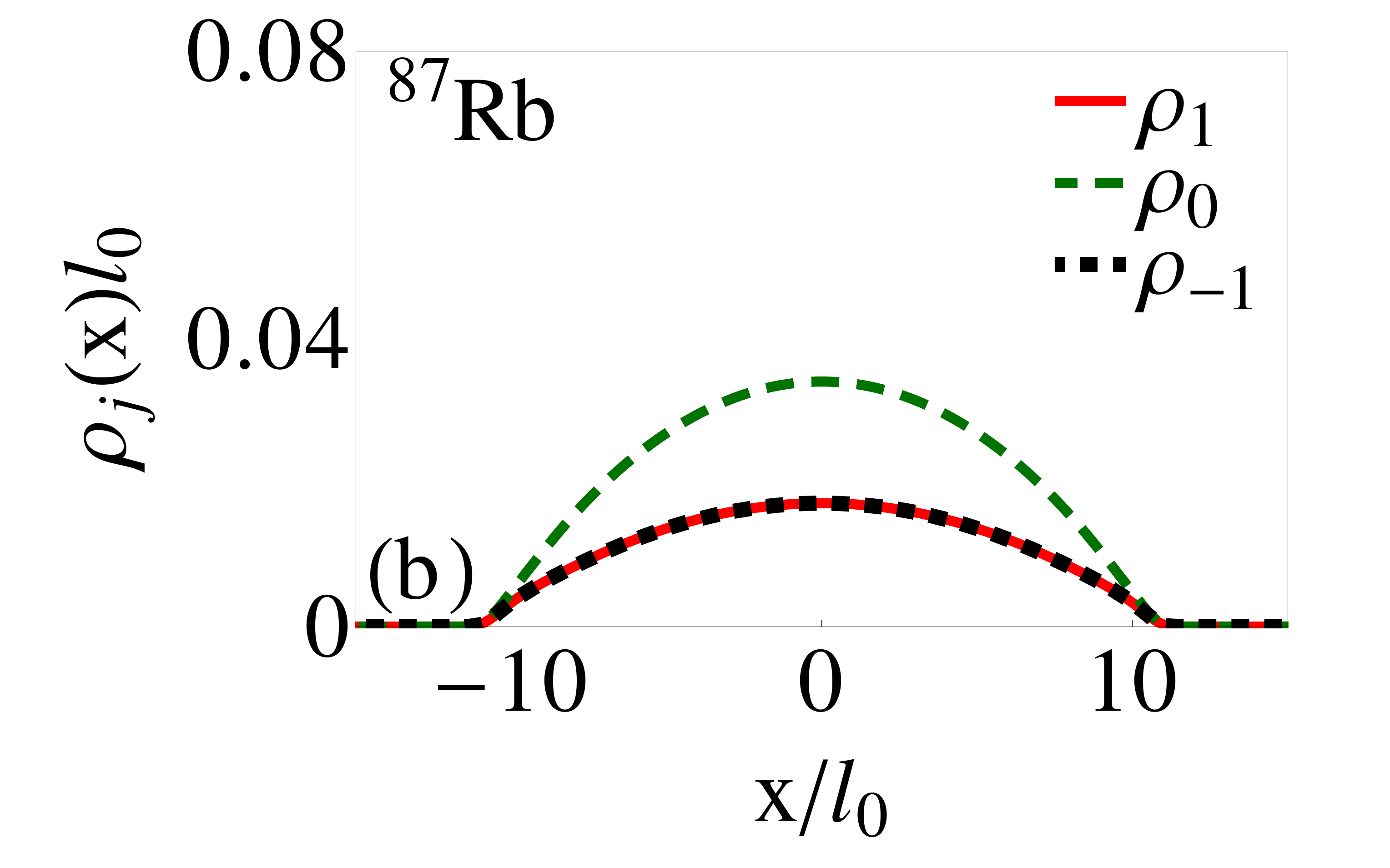}
\includegraphics[trim = 3mm 0mm 1cm 0mm, clip,width=.49\linewidth]{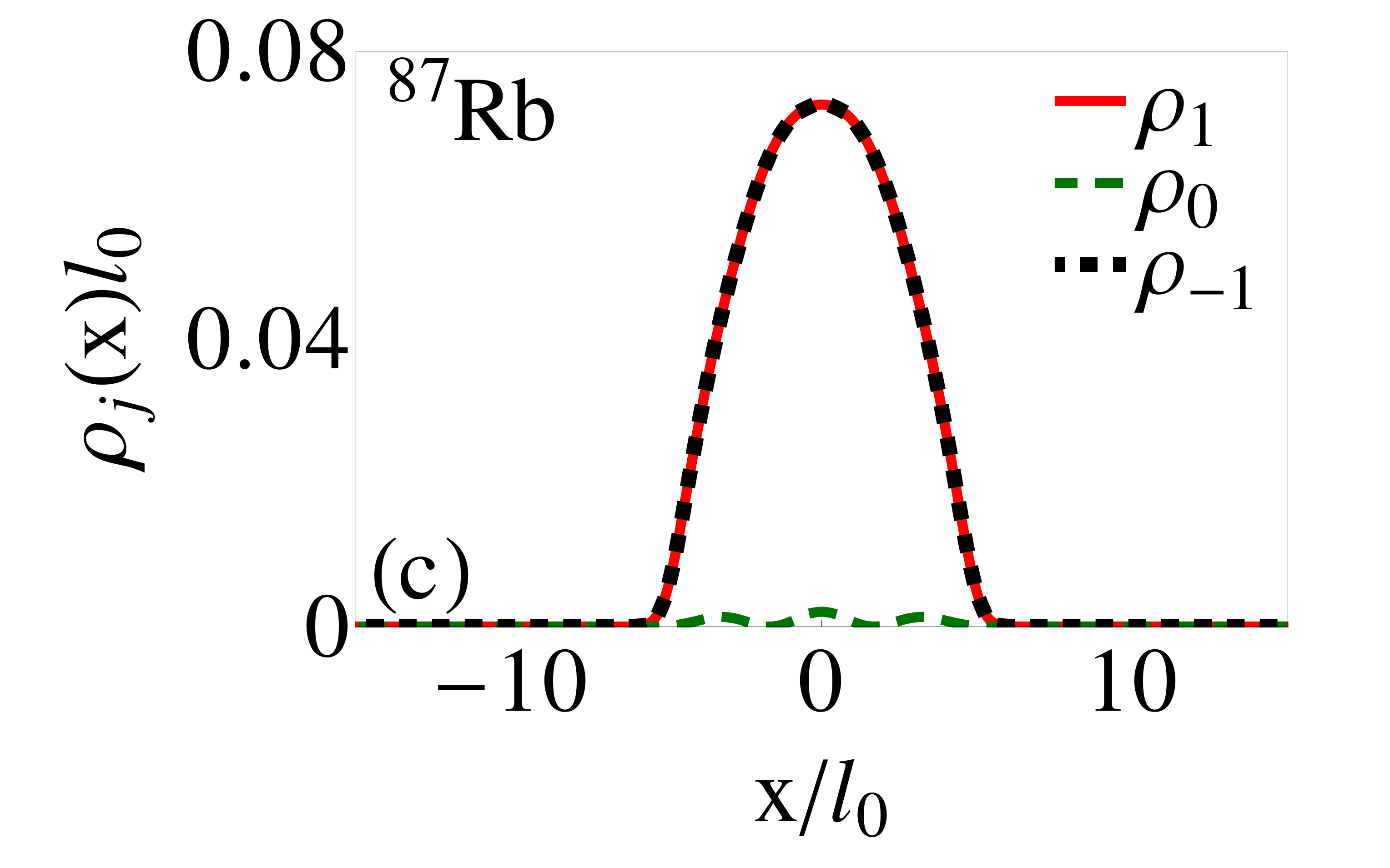}
\includegraphics[trim = 3mm 0mm 1cm 0mm, clip,width=.49\linewidth]{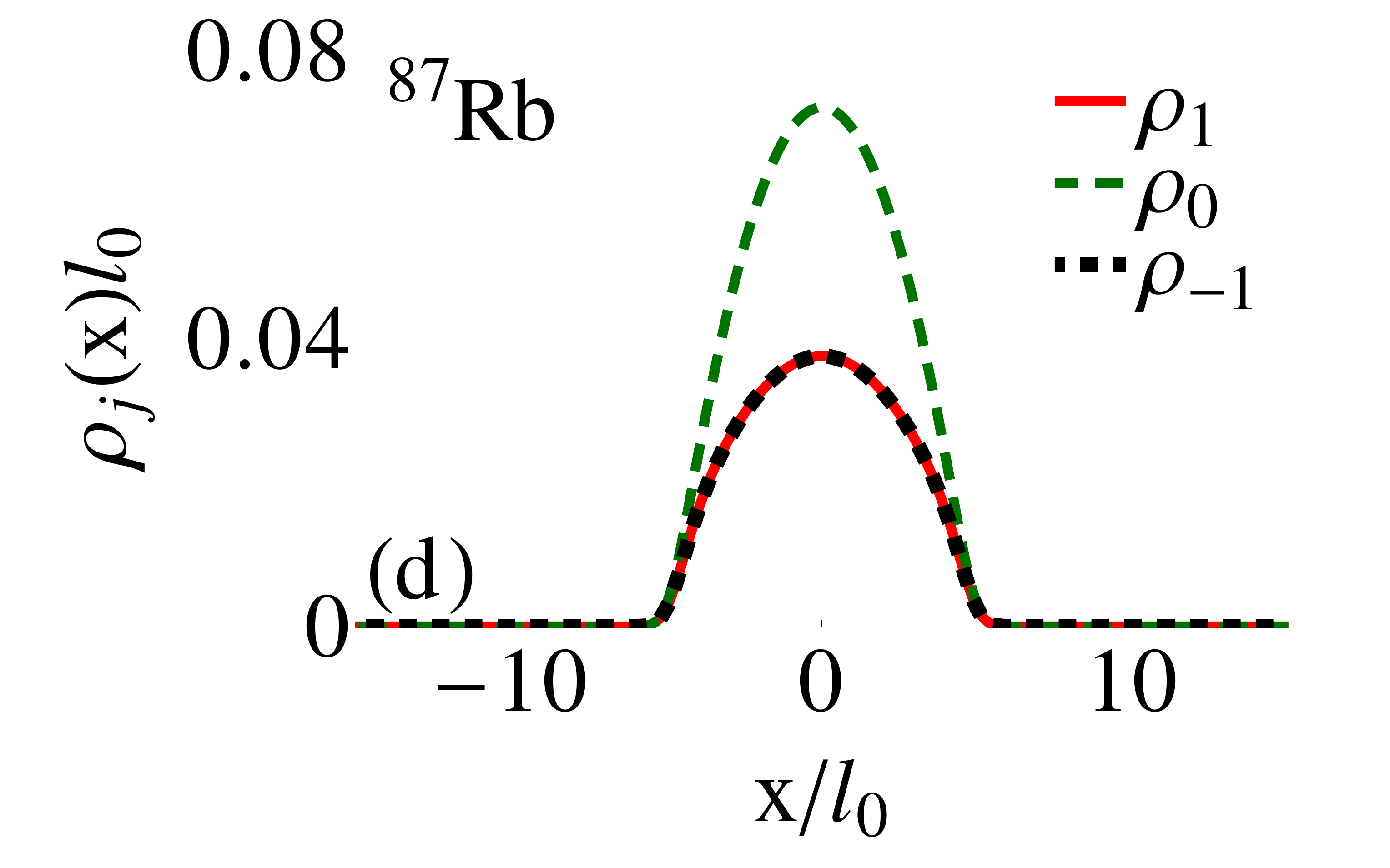}
\caption{(Color online) Ground state structure of $^{87}$Rb spinor BEC with 
$10^4$ atoms with (a) $N = 10000$, $\Omega=0.1$, $\gamma = 1$ 
(b) $N = 10000$, $\Omega = 1$, $\gamma = 1$
(c) $N = 1000$, $\Omega=0.1$, $\gamma = 1$, and  
(d) $N = 1000$, $\Omega=1$, $\gamma = 1$. Both the density and spatial coordinates 
are plotted in dimensionless units. Magnetization ${\cal M} = 0$
in all the cases.
}
\label{fig-3}\end{center}
\end{figure}

\begin{figure}[!b]
\begin{center}
\includegraphics[trim = 3mm 0mm 1cm 0mm, clip,width=.49\linewidth]{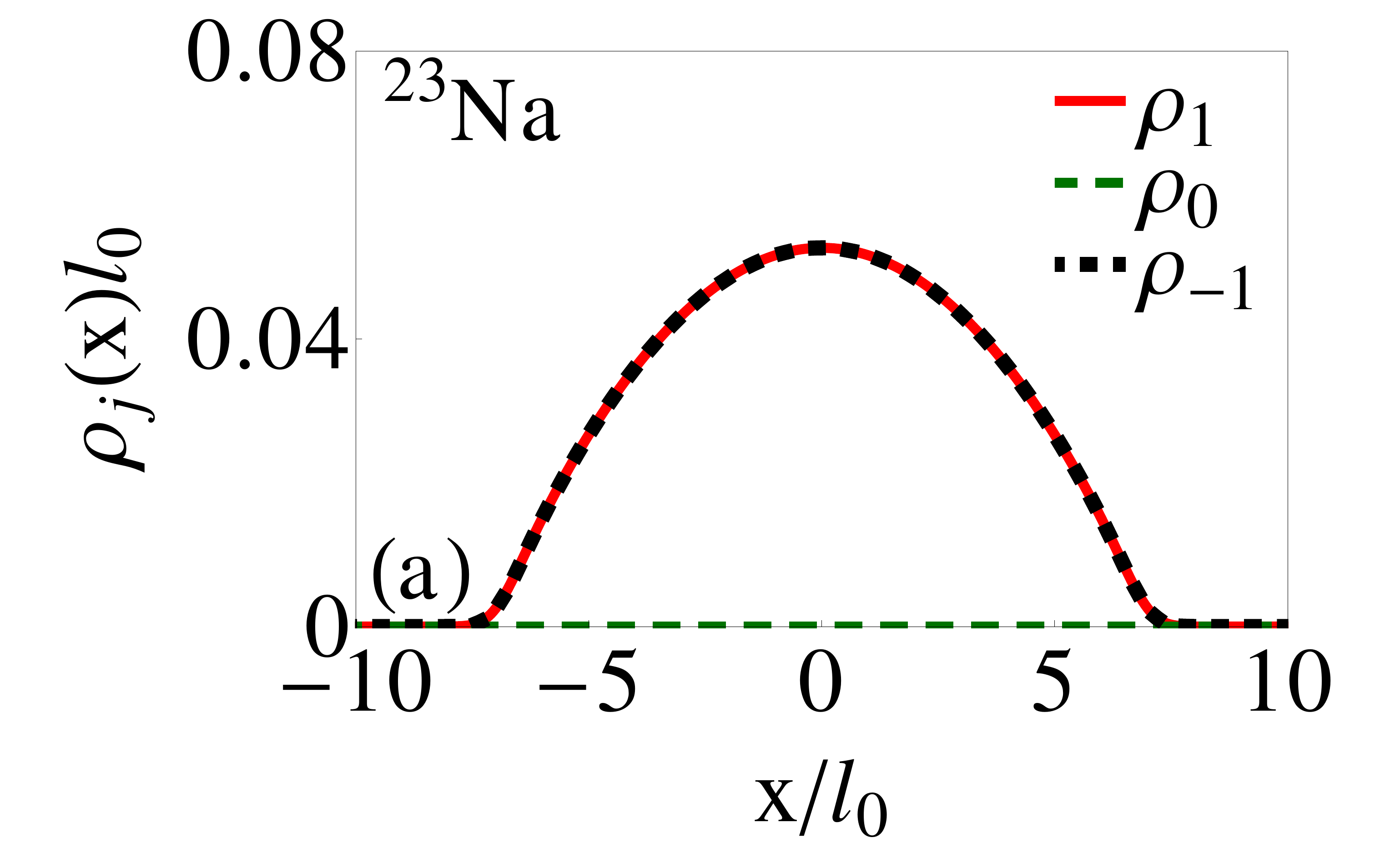}
\includegraphics[trim = 3mm 0mm 1cm 0mm, clip,width=.49\linewidth]{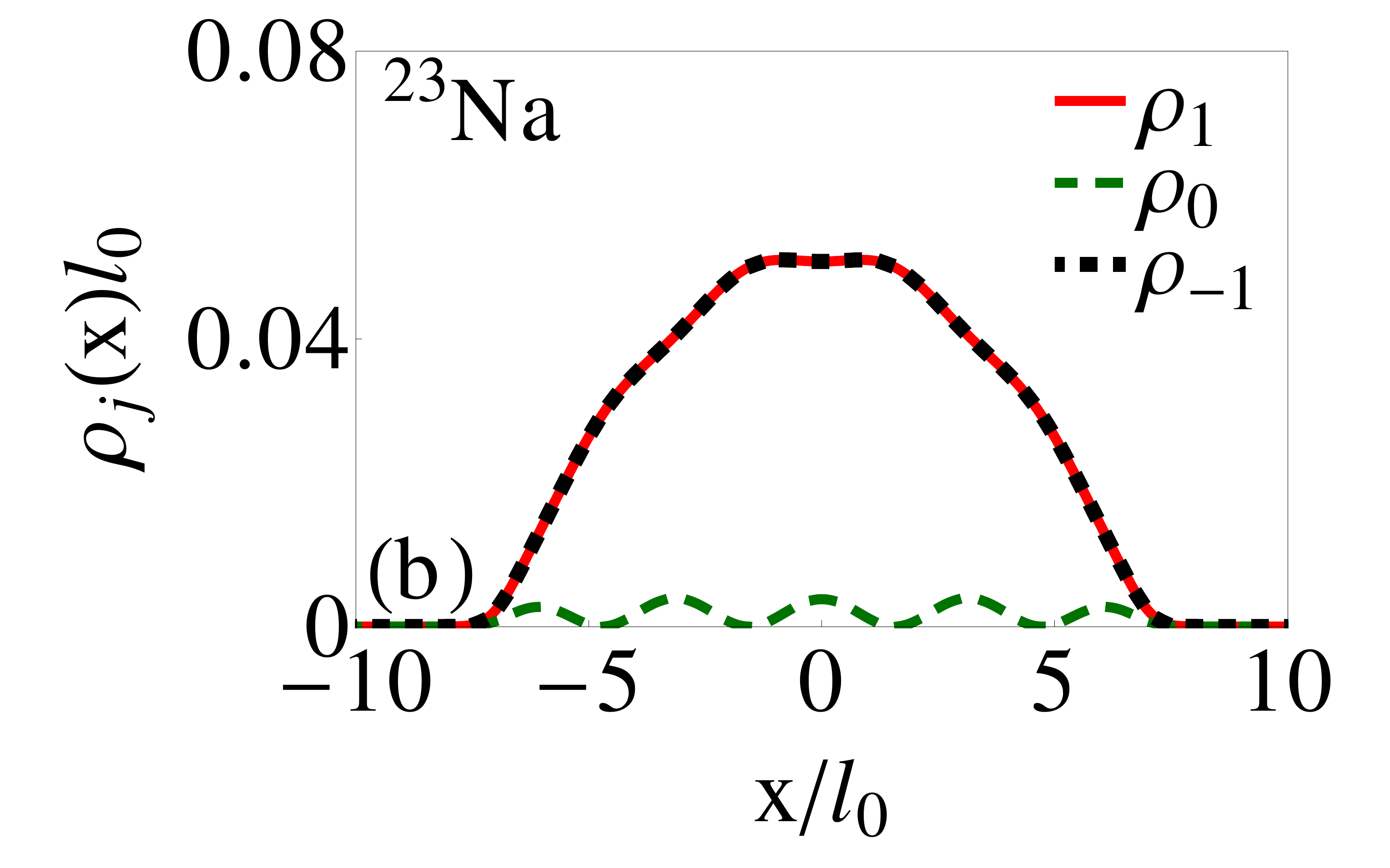}
\includegraphics[trim = 3mm 0mm 1cm 0mm, clip,width=.49\linewidth]{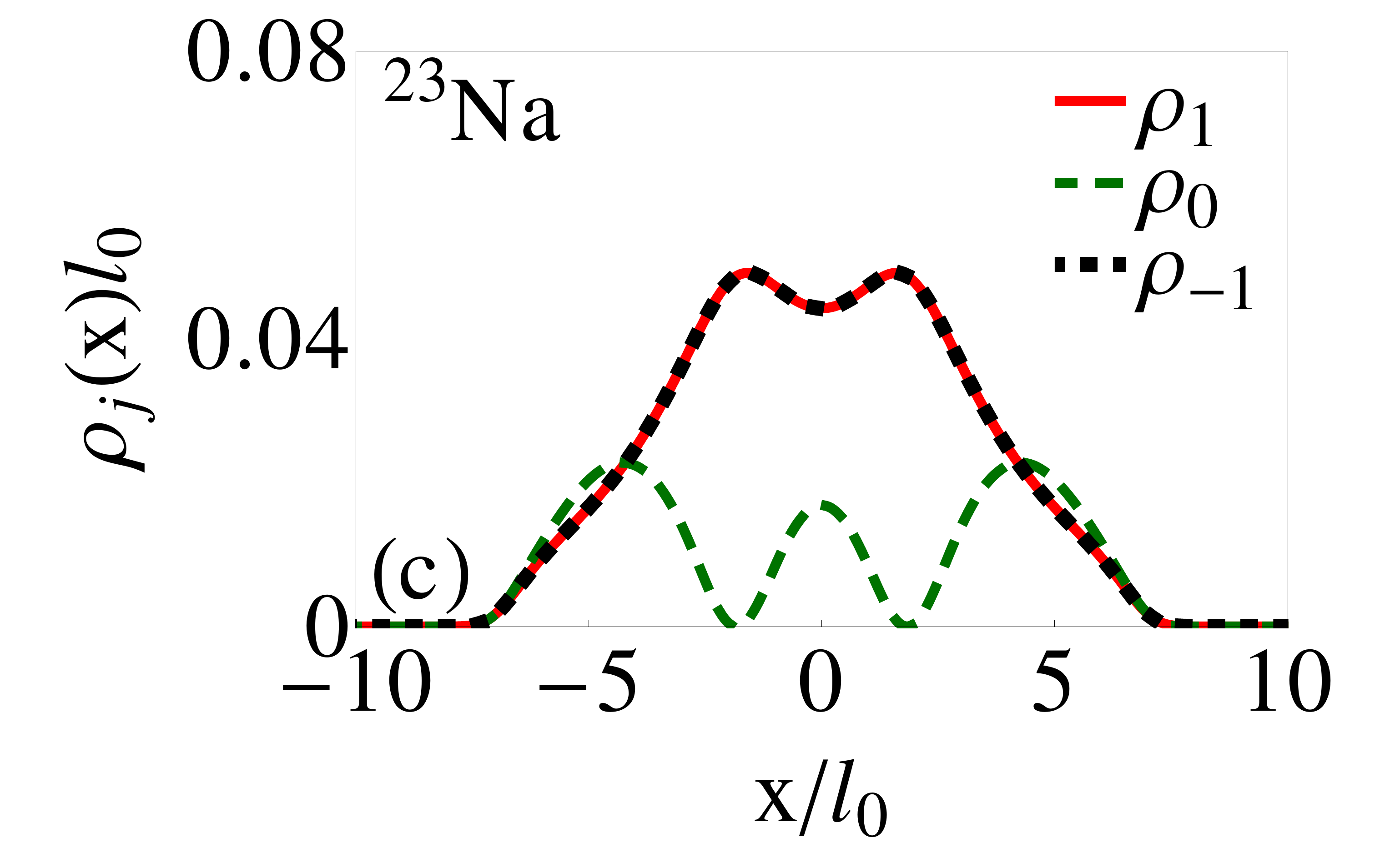}
\includegraphics[trim = 3mm 0mm 1cm 0mm, clip,width=.49\linewidth]{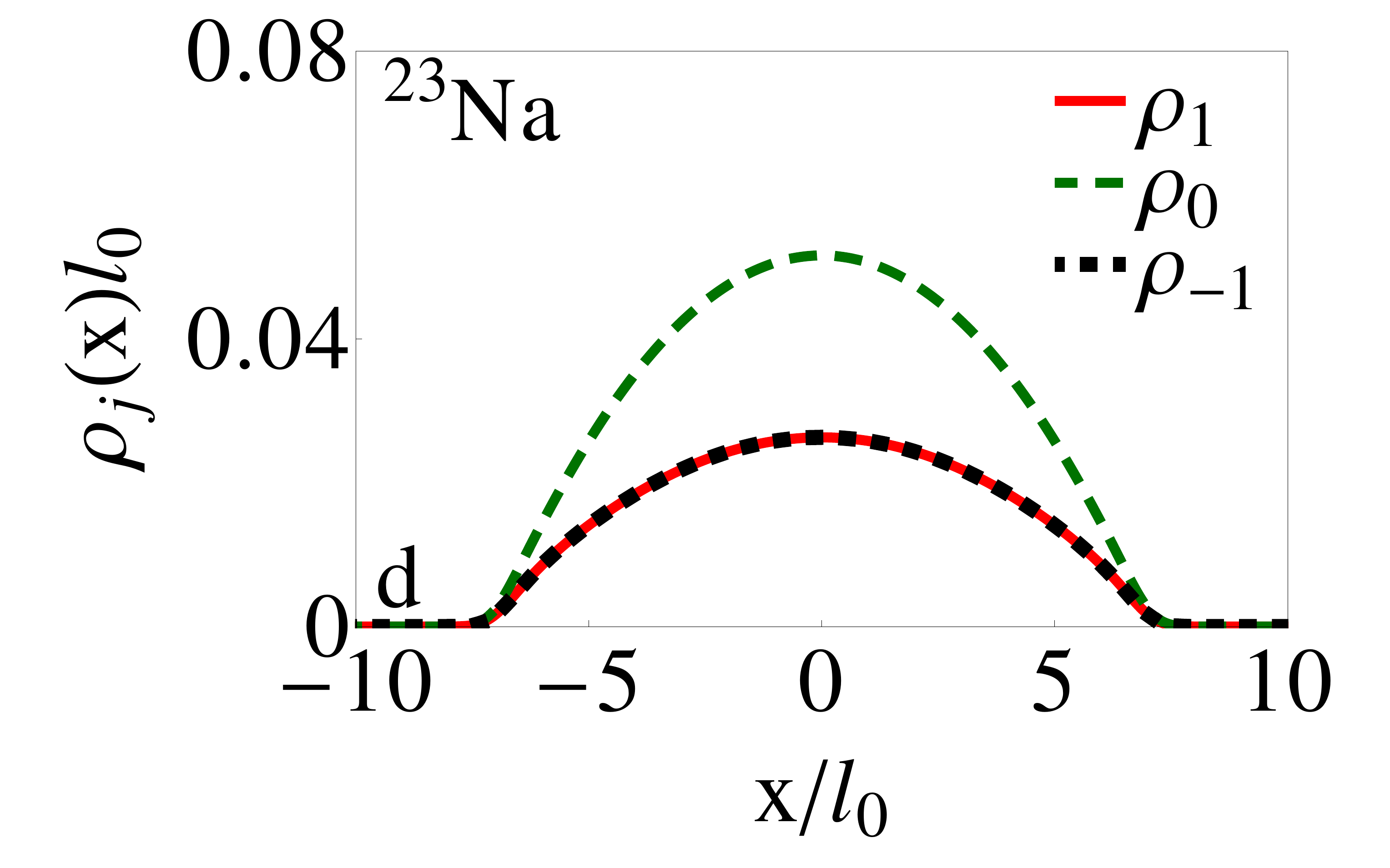}
 
\caption{(Color online) Ground state structure of $^{23}$Na spinor BEC with  
(a) $N = 10000$, $\Omega=0$, $\gamma = 1$, and also $\gamma=0$,
(b)  $N = 10000$, $\Omega = 0.5$, $\gamma = 1$, (c)  $N = 10000$, 
$\Omega = 1$, $\gamma = 1$, (d) $N = 10000$, $\Omega = 1.5$, 
$\gamma = 1$. Both the density and spatial coordinates are plotted in
dimensionless units. The magnetization ${\cal M}=0$ is zero in all the cases.
}
\label{fig-4}\end{center}
\end{figure}

We have also studied the effect of an increase in the   the Rabi 
term $\Omega$ on the state with $\rho_0(x)=0$ [viz. Fig. \ref{fig-1} (d)] 
maintaining magnetization ${\cal M}=0$. As discussed in Sec. \ref{Sec-III}, 
the Rabi term $\Omega$ favors miscibility of the system irrespective of the 
nature of the spin dependent interactions, while the  SO-coupling term $\gamma$ 
favors a phase separation. Hence, when both $\gamma$ and $\Omega$ are non zero, 
there is a competition between these two terms as one favors phase separation, 
whereas the other favors miscibility. To illustrate this, in Figs. \ref{fig-3} 
(a) and (b) we plot the component densities for $\Omega =0.1$ and $1$, 
respectively, for $N=10000$ and $\gamma =1$. The increase in the Rabi term 
$\Omega$ from $0.1$ to $1$ has transformed a phase-separated state to a 
miscible state. For smaller number of atoms, say $N = 1000$, we do not observe 
any phase separation with the increase in   the SO coupling 
$\gamma$. Nevertheless, the increase in $\gamma$ leads to a decrease in 
$\rho_0$ as is shown in Fig.~\ref{fig-3}(c), where $\rho_0$ is negligible in 
comparison to  overlapping $\rho_1$ and $\rho_{-1}$. Again as $\Omega$ is 
increased  in this case, the density $\rho_0$ first increases and ultimately 
ends up being larger than those of other two components 
[viz., Fig.~\ref{fig-3}(d)].

In the case of the SO-coupled polar BEC $^{23}$Na, we do not observe any 
phase separation consistent with the discussion of the uniform BEC in 
Sec. \ref{Sec-III}. In the absence of the Rabi term ($\Omega = 0$), the 
density profile in the presence and absence of the SO coupling are the same 
as is shown in Fig.~\ref{fig-4} (a). The introduction of the Rabi term leads 
to a non-zero density of the $0$th component as shown in Fig.~\ref{fig-4} (b) 
for $\Omega =0.5$. For both the ferromagnetic and polar BECs, in the presence 
of both SO coupling and Rabi terms, we observe a formation of structure in the 
ground state, where the $0$th component develops a train of dark notches as 
shown in Fig. \ref{fig-3}(c) for $^{87}$Rb and Figs. \ref{fig-4}(b) and (c) 
for $^{23}$Na. In $^{23}$Na, an increase in the Rabi term $\Omega$ leads to  
an increase in $\rho_0(x)$ from 0,  at the cost of $\rho_1(x)$ and 
$\rho_{-1}(x)$ as in the case of $^{87}$Rb, and ultimately, ends up with a 
solution where $\rho_0(x)>\rho_1(x)=\rho_{-1}(x)$.


\section{Summary }
\label{Sec-V}
We have studied the SO-coupled $F=1$ spinor BECs of $^{87}$Rb (ferromagnetic) 
and $^{23}$Na (antiferromagnetic or polar) atoms in quasi-1D traps. By 
comparing the energy of various competing structures for the SO-coupled spinor 
BEC in a 1D box, we have shown that any non-zero value of the SO coupling will 
lead to a phase separation between the $m_F=1$ and $m_F=-1$ components in the 
case of a ferromagnetic BEC in the absence of the Rabi term. On the other hand, 
for a polar BEC, SO coupling makes the miscible profile energetically more 
stable as compared to various possible phase-separated profiles. In the case 
of the trapped  SO-coupled BECs, we have numerically studied the ground state structures.
 In the ferromagnetic case, above a 
critical  number of atoms the BEC phase separates if the SO coupling strength 
exceeds a critical value in the absence of the Rabi term. The introduction of 
the Rabi term favors the miscibility for both the ferromagnetic and polar 
BECs. The present conclusions can be tested in experiments with present-day 
technology.


\begin{acknowledgements}
This work is financed by FAPESP (Brazil) under Contract No. 2013/07213-0 and 
also supported by CNPq (Brazil).
\end{acknowledgements}



\begin{thebibliography}{99}
\bibitem{Stamper-Kurn}
 D.~M.~Stamper-Kurn, M.~R.~Andrews, A.~P.~Chikkatur, S.~Inouye, H.-J.~Miesner, 
 J.~Stenger, and W.~Ketterle, 
 Phys. Rev. Lett. {\bf 80} 2027, (1998).
\bibitem{ueda}
 Y. Kawaguchi and M. Ueda, 
 Phys. Rep. {\bf 520}, 253 (2012).
\bibitem{luca}  
 L. Salasnich, A. Parola, and L. Reatto, 
 Phys. Rev. A {\bf 65}, 043614 (2002). 
\bibitem{Ohmi}
 T.~Ohmi, and K.~Machida, 
 J. Phys. Soc. Japan, {\bf 67}, 1822 (1998).
\bibitem{Ho}
 T.~L.~Ho, 
 Phys. Rev. Lett. {\bf 81}, 742 (1998).
\bibitem{young}
 J. Higbie and D. M. Stamper-Kurn, 
 Phys. Rev. Lett. {\bf 88}, 090401 (2002); 
 T. L. Ho and S. Zhang, Phys. Rev. Lett. {\bf 107}, 150403 (2011); 
 Y. Deng, J. Cheng, H. Jing, C. P. Sun, and S. Yi, 
 Phys. Rev. Lett. {\bf 108}, 125301 (2012); 
 J. Radic, T. A. Sedrakyan, I. B. Spielman, and V. Galitski,
 Phys. Rev. A {\bf 84}, 063604 (2011);.
\bibitem{Bychkov}
 Y.~A.~Bychkov and E.~I.~Rashba, 
 J. Phys. C {\bf 17}, 6039 (1984).
\bibitem{Dresselhaus}
 G.~Dresselhaus, 
 Phys. Rev. {\bf 100}, 580 (1955).
\bibitem{Liu}
 X.-J. Liu, M.~F.~Borunda, X.~Liu, and J.~Sinova,
 Phys. Rev. Lett. {\bf 102}, 046402 (2009).
\bibitem{Lin}
 Y.-J. Lin, K.~Jim\'enez-Garc\'ia, and I.~B.~Spielman, 
 Nature {\bf 471},  83 (2011).
\bibitem{Linx}
 V. Galitski and I. B. Spielman,
 Nature {\bf 494}, 49 (2013).
\bibitem{JY_Zhang}
 J.-Y. Zhang, S.-C. Ji, Z.~Chen, L.~Zhang, Z.-D. Du, B.~Yan, G.-S.~Pan, 
 B.~Zhao,
 Y.-J.~Deng, H.~Zhai, S.~Chen, and J.-W.~Pan,
 Phys. Rev. Lett. {\bf 109}, 115301 (2012);
 C.~Qu, C.~Hamner, M.~Gong, C.~Zhang, and P.~Engels,
 Phys. Rev. A {\bf 88}, 021604(R) (2013);
 M. Aidelsburger, M. Atala, and S. Nascimb´ene et al., 
 Phys. Rev. Lett. {\bf 107}, 255301 (2011);
 Z. Fu, P. Wang, and S. Chai, L. Huang, and J. Zhang,
 Phys. Rev. A {\bf 84}, 043609 (2011).
\bibitem{Juzeliunas}
 G.~Juzeli\={u}nas, J.~Ruseckas, and J.~Dalibard,
 Phys. Rev. A {\bf 81}, 053403 (2010);
 J. Dalibard {\em et al.}, 
 Rev. Mod. Phys. {\bf 83}, 1523 (2011).
\bibitem{P_Wang}
 P.~Wang, Z.-Q.~Yu, Z.~Fu, J.~Miao, L.~Huang,
 S.~Chai, H.~Zhai, and J.~Zhang,
 Phys. Rev. Lett. {\bf 109}, 095301 (2012);
 L.~W.~Cheuk, A.~T.~Sommer, Z.~Hadzibabic, T.~Yefsah, W.~S.~Bakr, 
 and M.~W.~Zwierlein,
 Phys. Rev. Lett. {\bf 109}, 095302 (2012).
\bibitem{H_Wang}
 H.~Wang,
 Int. J. of Computer Math.
 {\bf 84}, 925 (2007).
\bibitem{Bao}
  W.~Bao and F.~Y.~Lim,
  Siam J. Sci. Comp. {\bf 30}, 1925 (2008);
  F.~Y.~Lim and W.~Bao,
  Phys. Rev. E {\bf 78}, 066704 (2008).
\bibitem{Wang}
 C.~Wang, C.~Gao, C-M Jian, and H.~Zhai,
 Phys. Rev. Lett. {\bf 105}, 160403 (2010); A.
 Aftalion and P. Mason, Phys. Rev. A 88, 023610 (2013);
 R. Gupta, G. S. Singh, and J. Bosse,
 Phys. Rev. A {\bf 88}, 053607 (2013);
 Q.-Q. Lu and D. E. Sheehy,
 Phys. Rev. A {\bf 88}, 043645 (2013).
\bibitem{Gopalakrishnan}
 T.~D.~Stanescu, B.~Anderson, and V.~Galitski,
 Phys. Rev. A {\bf 78}, 023616 (2008);
 C.-J Wu and I.~Mondragon-Shem, X.-F. Zhou
 Chin. Phys. Lett. {\bf 28}, 097102 (2011);
 Q. Zhou and X. Cui,
 Phys. Rev. Lett. {\bf 110}, 140407 (2013);
 S.~Gopalakrishnan, A.~Lamacraft, and P.~M.~Goldbart,
 Phys. Rev. A {\bf 84}, 061604(R) (2011);
 H.~Hu, B.~Ramachandhran, H.~Pu, and X.-J. Liu,
 Phys. Rev. Lett. {\bf 108}, 010402 (2012);
 B.~Ramachandhran, B.~Opanchuk, X.-J.~Liu, H.~Pu, P.~D.~Drummond, and H.~Hu,
 Phys. Rev. A {\bf 85}, 023606 (2012);
 S.~Sinha, R.~Nath, and L.~Santos, 
 Phys. Rev. Lett. {\bf 107}, 270401 (2011);
 T.~Ozawa and G.~Baym, 
 Phys. Rev. A {\bf 85}, 013612 (2012).
\bibitem{Xu-1}
 Z.~F.~Xu, Y.~Kawaguchi, L.~You, and M.~Ueda, 
 Phys. Rev. A {\bf 86}, 033628 (2012); 
 S.-W. Song, Y.-C. Zhang, H. Zhao, X. Wang, and W.-M. Liu,
 Phys. Rev. A {\bf 89}, 063613 (2014);
 P.-S. He, Y.-H. Zhu, and W.-M. Liu
 Phys. Rev. A {\bf 89}, 053615 (2014);
 Y. Deng, J. Cheng, H. Jing, and S. Yi,
 Phys. Rev. Lett. {\bf 112}, 143007 (2014);
 K. Riedl, C. Drukier, P. Zalom, and P. Kopietz,
 Phys. Rev. A {\bf 87}, 063626 (2013);
 T.~Kawakami, T.~Mizushima, and K.~Machida, 
 Phys. Rev. A {\bf 84}, 011607 (2011);
 Z.~F.~Xu, R.~L\"{u}, and L.~You, 
 Phys. Rev. A {\bf 83}, 053602 (2011);
 S.-K. Yip, 
 Phys. Rev. A {\bf 83}, 043616 (2011);
 S.-W.~Su, I.-K.~Liu, Y.-C.~Tsai, W.~M.~Liu, and S.-C.~Gou, 
 Phys. Rev. A {\bf 86}, 023601 (2012);
 Y.~Zhang, L.~Mao, and C.~Zhang,
 Phys. Rev. Lett. {\bf 108}, 035302 (2012);
 S.-W.~Song, Y.-C. Zhang, L.~Wen, and H.~Wang,
 J. Phys. B {\bf 46}, 145304 (2013).
\bibitem{Ruokokoski}
 E.~Ruokokoski, J.~A.~M.~Huhtam\"{a}ki, and M.~M\"{o}tt\"{o}nen,
 Phys. Rev A {\bf 86}, 051607(R) (2012).
\bibitem{Larson}
 J.~Larson, J.-P.~Martikainen, A.~Collin, and E.~Sj\"{o}qvist,
 Phys. Rev. A 82, 043620 (2010).
\bibitem{Merkl}
 M. Merkl, A. Jacob, F. E. Zimmer, P. \"{O}hberg, and L. Santos,
 Phys. Rev. Lett. {\bf 104}, 073603 (2010).
\bibitem{super} 
 T. Ozawa, L. P. Pitaevskii, and S. Stringari,
 Phys. Rev. A {\bf 87}, 063610 (2013);
 D. W. Zhang, J. P. Chen, C. J. Shan, Z. D. Wang, and
 S. L. Zhu, Phys. Rev. A {\bf 88}, 013612 (2013); 
 Q. Zhu, C. Zhang and B. Wu, 
 Europhys. Lett. {\bf 100}, 50003 (2012);
 D. Toniolo and J. Linder,
 Phys. Rev. A {\bf 89}, 061605(R) (2014).
\bibitem{josep}
 M. A. Garcia-March, G. Mazzarella, L. Dell'Anna, 
 B. Juli\'a-D\'iaz, L. Salasnich, and A. Polls,
 Phys. Rev. A {\bf 89}, 063607 (2014).
\bibitem{vor} 
 A. L. Fetter,
 Phys. Rev. A {\bf 89}, 023629 (2014).
 X. F. Zhou, J. Zhou, and C. Wu, 
 Phys. Rev. A {\bf 84}, 063624 (2011);
 Z.-F. Xu, S. Kobayashi, and M. Ueda,
 Phys. Rev. A {\bf 88}, 013621 (2013);
 C.-F. Liu, Y.-M. Yu, S.-C. Gou, and W.-M. Liu,
 Phys. Rev. A {\bf 87}, 063630 (2013).
\bibitem{sol}
 H. Sakaguchi, Ben Li, and B. A. Malomed,
 Phys. Rev. E {\bf 89}, 032920 (2014);
 Y. Xu, Y. Zhang, and B. Wu, Phys. 
 Rev. A {\bf 87}, 013614 (2013); 
 O. Fialko, J. Brand, and U. Z¨ulicke, 
 Phys. Rev. A {\bf 85}, 051605(R) (2012).
\bibitem{Zhou}
 F. Zhou,
 Phys. Rev. Lett. 87, 080401 (2001);
 S.~Yi, \"{O}. E. M\"{u}stecaplioglu, C.~P.~Sun, and L.~You,
 Phys. Rev. A {\bf 66}, 011601(R) (2002);
 W.~Zhang, S.~Yi, and L.~You,
 New J. Phys. {\bf 5}, 77 (2003);
 K.~Murata, H.~Saito, and M.~Ueda,
 Phys. Rev. A {\bf 75}, 013607 (2007).
\bibitem{Matuszewski-1}
 M.~Matuszewski, T.~J.~Alexander, and Y.~S.~Kivshar,
 Phys. Rev. A 80, 023602 (2009).
\bibitem{Matuszewski-2}
 M.~Matuszewski,
 Phys. Rev. A {\bf 82}, 053630 (2010).
\bibitem{Ao}
 P.~Ao and S.~T.~Chui,
 Phys. Rev. A {\bf 58}, 4836 (1998);
 P. Facchi, G. Florio, S. Pascazio, and F. V. Pepe,
 J. Phys. A: Math. Theor. {\bf 44} 505305 (2011).


\bibitem{lan} Z. Lan and P. \"Ohberg, \pra {\bf 89,} 023630 (2014). 

\bibitem{Y_Zhang}
 Y. Li, G. I. Martone, L. P. Pitaevskii, and S. Stringari,
 Phys. Rev. Lett. {\bf 110}, 235302 (2013);
 Y.~Zhang and C.~Zhang,
 Phys. Rev. A {\bf 87}, 023611 (2013);
 L.~Salasnich and B.~A.~Malomed,
 Phys. Rev. A {\bf 87}, 063625 (2013);
D. A. Zezyulin, R. Driben, V. V Konotop, and B. A. Malomed,
 Phys. Rev. A {\bf 88}, 013607  (2013);
 Y. Cheng, G. Tang, and S. K. Adhikari,
 Phys. Rev. A {\bf 89}, 063602 (2014). 
\bibitem{Gautam}
 S. Gautam and D. Angom,
 J. Phys. B {\bf 44}, 025302 (2011);
 S. Gautam and D. Angom,
 J. Phys. B {\bf 43}, 095302 (2010).
\bibitem{Isoshima}
 T.~Isoshima, K.~Machida, and T.~Ohmi,
 J. Phys. Soc.  Japan {\bf 70}, 1604 (2001).
\bibitem{Muruganandam}
 P.~Muruganandam and S.~K.~Adhikari,
 Comput. Phys. Commun. {\bf 180}, 1888 (2009);
 D. Vudragovic, I. Vidanovic, A. Balaz, P. Muruganandam, and S. K. Adhikari.
 Comput. Phys. Commun. {\bf 183}, 2021 (2012). 



\end{thebibliography}
\end{document}